\documentclass[fleqn,usenatbib]{mnras}

\usepackage{newtxtext,newtxmath}

\usepackage[T1]{fontenc}

\usepackage{multirow}
\usepackage{multicol}
\usepackage{makecell}
\usepackage{graphicx}
\usepackage[flushleft]{threeparttable}
\usepackage{booktabs}
\usepackage{comment}
\usepackage{hyperref}
\DeclareRobustCommand{\VAN}[3]{#2}
\let\VANthebibliography\thebibliography
\def\thebibliography{\DeclareRobustCommand{\VAN}[3]{##3}\VANthebibliography}

\usepackage{graphicx}	
\usepackage{amsmath}	
\usepackage{lscape} 
\usepackage{multirow}
\usepackage{longtable}
\usepackage{chemfig}

\newcommand{\kjm}{\,kJ mol$^{-1}$}	

\title[HC$_5$N gas-phase formation network] {A comprehensive study of the gas-phase formation network of HC$_5$N: theory, experiments, observations and models}

\author[Giani et al.]{
Lisa Giani,$^{1,2}$\thanks{E-mail:lisa.giani@univ-grenoble-alpes.fr}
Eleonora Bianchi,$^{3,4}$
Martin Fournier,$^{5,6}$
Sidaty Cheikh Sid Ely,$^{5}$
Cecilia Ceccarelli,$^{1}$\thanks{E-mail:cecilia ceccarelli@univ-grenoble-alpes.fr}
\and
Marzio Rosi,$^{7}$
Jean-Claude Guillemin,$^{8}$
Ian R. Sims$^{5,9}$\thanks{E-mail:ian.sims@univ-rennes.fr}
and Nadia Balucani$^{1,2}$\thanks{E-mail:nadia.balucani@unipg.it}
\\\\
$^{1}$Univ. Grenoble Alpes, CNRS, IPAG, 38000 Grenoble, France\\
$^{2}$Dipartimento di Chimica, Biologia e Biotecnologie, Universita degli Studi di Perugia, Perugia, 06123, Italy\\
$^{3}$ INAF, Osservatorio Astrofisico di Arcetri, Largo E. Fermi 5, I-50125, Firenze, Italy\\
$^{4}$ Excellence Cluster ORIGINS, Boltzmannstraße 2, 85748 Garching bei München, Germany\\
$^{5}$Univ Rennes, CNRS, IPR (Institut de Physique de Rennes) - UMR 6251, F-35000 Rennes, France\\
$^{6}$Institute of Chemical Sciences, School of Engineering and Physical Sciences, Heriot-Watt University, Edinburgh, EH14 4AS, UK\\
$^{7}$Dipartimento di Ingegneria Civile e Ambientale, Universita degli Studi di Perugia, Perugia, 06125, Italy\\
$^{8}$Univ Rennes, École Nationale Supérieure de Chimie de Rennes, CNRS, ISCR - UMR 6226, F-35000 Rennes, France\\
$^{9}$Institut universitaire de France (IUF)\\
}

\date{Accepted XXX. Received YYY; in original form ZZZ}

\pubyear{2023}

\begin{document}
\label{firstpage}
\pagerange{\pageref{firstpage}--\pageref{lastpage}}
\maketitle

\begin{abstract}

Cyanopolyynes are among the largest and most commonly observed interstellar Complex Organic Molecules in star-forming regions. 
They are believed to form primarily in the gas-phase, but their formation routes are not well understood.
We present a comprehensive study of the gas-phase formation network of cyanobutadiyne, HC$_5$N, based on new theoretical calculations, kinetics experiments, astronomical observations, and astrochemical modeling.
We performed new quantum mechanics calculations for six neutral-neutral reactions in order to derive reliable rate coefficients and product branching fractions. 
We also present new CRESU data on the rate coefficients of three of these reactions (C$_3$N + C$_2$H$_2$, C$_2$H + HC$_3$N, CN + C$_4$H$_2$) obtained at temperatures as low as 24 K. 
In practice, six out of nine reactions currently used in astrochemical models have been updated in our reviewed network.
We also report the tentative detection of the $^{13}$C isotopologues of HC$_5$N in the L1544 prestellar core. 
We derived a lower limit of $^{12}$C/$^{13}$C > 75 for the HC$_5$N isotopologues, which does not allow to bring new constraints to the HC$_5$N chemistry. 
Finally, we verified the impact of the revised reactions by running the GRETOBAPE astrochemical model. 
We found good agreement between the HC$_5$N predicted and observed abundances in cold ($\sim$10 K) objects, demonstrating that HC$_5$N is mainly formed by neutral-neutral reactions in these environments.
In warm molecular shocks, instead, the predicted abundances are a factor of ten lower with respect to observed ones.
In this environment possessing an higher gas ionization fraction, we speculate that the contribution of ion-neutral reactions could be significant. 

\end{abstract}




\section{Introduction} \label{sec:introduction}

So far, more than 330 molecules have been detected in the interstellar medium (ISM), and this number is steadily increasing in time with the advent of more sensitive facilities \citep[e.g.,][and \url{https://cdms.astro.uni-koeln.de/classic/molecules}]{McGuire2022-ISMcensus}.
Approximately 40\% of the detected interstellar species are organic and contain at least six atoms; these are known as interstellar Complex Organic Molecules \citep[iCOMs or COMs;][]{herbst2009complex, ceccarelli2017seeds}. Many, though not all, are found in star-forming regions.
A significant portion of them are nitriles, that is, organic molecules holding a cyano group (C$\equiv$N).

Molecules belonging to the family of cyanopolyynes, from cyanoacetylene (HC$_3$N) to cyanodecapentayne (HC$_{11}$N), are among the largest detected iCOMs, and they hold particular importance for several reasons. 
First, they are almost ubiquitous in the ISM and, more important in the context of this work, are abundant in star-forming regions, especially during the cold molecular and prestellar phases \citep[e.g.,][]{1971Turner, 1976Avery, 1977Kroto, 1978Broten, 1992KawaguchiA, 1992Suzuki, 2008Sakai, rolffs2011structure,2012Chapillon, aljaber2017history, calcutt2018almanitriles,loomis2021investigation, mondal2023investigating, bianchi2023cyanopolyyne}.
Second,  given their resistance to destruction from the interstellar UV photons that permeate the Milky Way \citep{1995clarkeferris, silva2009uv, leach2014ionization}, cyanopolyynes may provide an important contribution to the possible inheritance of carbon and molecular complexity during the various phases of the formation of planetary systems \citep[e.g.,][]{bockelee2000new, vuitton2007ion,  mumma2011chemical, oberg2015comet}. 
Third, cyanopolyynes have also been detected in the comet 67P/Churyumov-Gerasimenko by the Rosetta mission, suggesting that either they are formed in situ or are inherited by the Proto Solar Nebula \citep{rubin2019, 2021Hanni}.
Finally, the family head, cyanoacetylene, is a precursor of amino acids \citep{Sanchez1966-HCNlife,2015Patel}, nucleobases \citep{ferris1968studies, crowe2006reaction,choe2021can}, ribonucleotides \citep{ingar2003synthesis, powner2009synthesis} and, thus, a possibly important prebiotic molecule. 
While there is less information regarding the prebiotic role of larger cyanopolyynes, it is likely that they can also hydrolyze in the presence of water or lead to the formation of tholins  \citep{kaiser2012formation,pentsak2024role}.
For these reasons, it is particularly important to fully understand how cyanopolyynes are formed and destroyed. 

From a chemical point of view, cyanopolyynes are a family of highly unsaturated nitriles characterized by the presence of a cyano group at the end of a polyyne chain. Their general formula is HC$_n$N, where $n$ takes on odd values (5, 7, ...), while even values (4, 6, ...) are associated with related polyynes (C$_n$H$_2$). 
The first question to answer is, therefore, whether the cyanopolyynes growth depends on reactions involving smaller cyanopolyynes and/or polyynes.
In general, molecules can be formed in the gas phase or on the interstellar grain surfaces \citep[see, e.g.,][for a recent review on large molecules]{Ceccarelli2023-PP7}.
However, cyanopolyynes are likely formed in the gas phase for two main reasons: (i) on the icy surfaces of dust grains, they would readily react with atomic hydrogen, leading to the formation of fully saturated counterparts; and (ii) if they were formed on the grain surfaces, there is no clear mechanism for releasing them into the gas phase in the cold environments where they have also been detected. Therefore, the focus of this work will be on the gas-phase formation of cyanopolyynes.

In the literature, two classes of gas-phase reactions have been considered to account for the formation of cyanopolyynes \citep[see, e.g., the recent review on the subject by][]{bianchi2023cyanopolyyne}: 
ion-neutral \citep[e.g.,][]{herbst1989gas} and neutral-neutral reactions \citep[e.g.,][]{herbst1990gas,chastaing1998neutral,huang2000crossed,zhang2009crossed,loison2013gas}:
\begin{enumerate}
    \item C$_{2n}$H$_{2}$ + CN $\rightarrow$ HC$_{2n+1}$N + H
    \item C$_{n+1}$N + C$_{2}$H$_{2}$  $\rightarrow$  HC$_{2n+1}$N + H
    \item HC$_{n+1}$N + C$_2$H $\rightarrow$ HC$_{2n+1}$N + H
    \item C$_{2n+2}$H + N $\rightarrow$ HC$_{2n+1}$N + C
    \item C$_{2n+1}$H$_2$ + N $\rightarrow$ HC$_{2n+1}$N + H
    \item C$_{2n}$H + HCN or HNC $\rightarrow$  HC$_{2n+1}$N + H
    \item HC$_{2n+1}$NH$^+$ + $e^-$ $\rightarrow$  HC$_{2n+1}$N + H
    \item H$_3$C$_{2n+1}$N$^+$ + $e^-$ $\rightarrow$  HC$_{2n+1}$N + H$_2$
\end{enumerate}

To conduct a comprehensive study with a network of reliable formation routes, we have limited this study to the case of HC$_5$N (cyanodiacetylene, also known as cyanobutadiyne or 2,4-pentadiynenitrile)
by combining new quantum chemistry calculations, experimental results, astronomical observations, and astrochemical modeling.

The article is organized as follows. 
In Sec.~\ref{sec:overview-gas-routes}, we present a critical review of what was known about the HC$_5$N gas-phase formation reactions before our work.
The employed experimental and theoretical methods are described in Sec.~\ref{sec:methods}. 
The experimental and theoretical results are described in Sec.~\ref{sec:experiments-results} and \ref{sec:theo-results} (more detailed information is given in the Appendix, sections \ref{sec:appendix-results} and \ref{sec:appendix-results-theo}).
Section~\ref{sec:observations} reports on the astronomical observations.
The upgrade of the network reactions, based on our new data on the reactions of interest, is summarised in Sec.~\ref{sec:revised-network}, and the astrochemical model results obtained with the new network are shown in Sec.~\ref{sec:astro-modeling}.
In Sec.~\ref{sec:discussion}, we discuss the results by comparing the model predictions and the observations. 
Finally, the conclusions are presented in Sec.~\ref{sec:conclusions}.


\section{Critical overview of the gas-phase reactions}\label{sec:overview-gas-routes}

\begin{figure}
    \includegraphics[width=8.5cm]{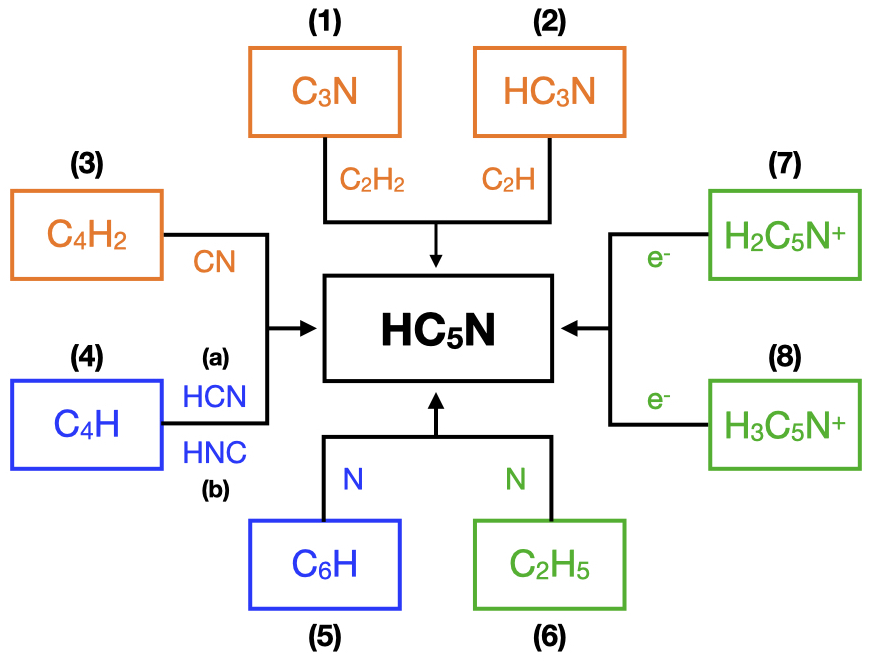}
    \caption{Scheme of the gas-phase reactions involved in the formation of HC$_5$N and listed in Tab.~\ref{tab:revised-network}. The numbers on top of each box are those in Tab.~\ref{tab:revised-network} and used in the subsections of Sec.~\ref{sec:overview-gas-routes}. The colour code of the boxes is the following: Orange, reactions that have been studied both experimentally and theoretically in this work; Blue, reactions for which we have carried out QM calculations and no experimental data are available; Green, reactions not revised in this work and assumed as correctly reported in the databases.}
    \label{fig:reac-scheme}
\end{figure}

We carried out a literature search, starting from the list of reactions reported in the astrochemical databases KIDA \citep[Kinetic Database for Astrochemistry:][]{wakelam2012kinetic,wakelam2024} and UDfA \citep[UMIST Database for Astrochemistry:][]{mcelroy2013umist,Umist2022}.
The neutral-neutral and ion-neutral reactions that have been proposed to explain the formation of HC$_5$N are (see Fig. \ref{fig:reac-scheme}):
\begin{enumerate}
    \item[1.] C${_3}$N + C$_{2}$H$_{2}$  $\rightarrow$  HC$_{5}$N + H
    \item[2.] HC${_3}$N + C$_2$H $\rightarrow$ HC$_{5}$N + H
    \item[3.] C$_{4}$H$_{2}$ + CN $\rightarrow$ HC$_{5}$N + H
    \item[4.] C$_{4}$H + HCN or HNC $\rightarrow$  HC$_{5}$N + H
    \item[5.] C${_6}$H + N $\rightarrow$ HC$_{5}$N + C
    \item[6.] C${_5}$H$_2$ + N $\rightarrow$ HC$_{5}$N + H
    \item[7.] HC$_{5}$NH$^+$ + $e^-$ $\rightarrow$  HC$_{5}$N + H
    \item[8.] H$_3$C$_{5}$N$^+$ + $e^-$ $\rightarrow$  HC$_{5}$N + H$_2$
\end{enumerate}

A review of what is known for each reaction is reported in the following sections. 

\subsection{Reaction [1]: \texorpdfstring{C${_3}$N + C$_{2}$H$_{2}$  $\rightarrow$  HC$_{5}$N + H}{reaction1} \label{subsec:overview-reaction1}}

This reaction is included in KIDA with an estimated rate coefficient of $k(T)=1.3\times10^{-10}$ cm$^3$ s$^{-1}$ in the 50-200 K range of temperatures. Another value of $k(T)=2.19\times10^{-10}$ cm$^3$ s$^{-1}$ is associated with the C$_2$H + HC$_3$N channel. In the last release 
of the UDfA \citep{Umist2022}, the reaction is included with the experimental rate coefficient obtained with the CRESU technique and that we are reporting for the first time in this manuscript. 

There are two possible exothermic channels: one leading to HC$_5$N + H ($\Delta H$ $\sim -$ 140 kJ mol$^{-1}$) and another one (less exothermic with $\Delta H$ $\sim -$ 10 kJ mol$^{-1}$) leading to HC$_3$N + C$_2$H. To the best of our knowledge, neither the potential energy surface (PES) nor the product branching fractions (BFs) have been ever reported. Thus, we carried out dedicated QM calculations to characterize the PES and determine the main reaction products via RRKM estimates. 

Our experimental results are described in Sec.~\ref{subsubsec:experiments-results-reaction1} and in the Appendix \ref{subsec:append-exp-R1}, while our theoretical results, which include the product branching fractions, are reported in Sec.~\ref{subsubsec:theo-results-reaction1} and in the Appendix \ref{subsec:append-theo-R1}.

\subsection{Reaction [2]: \texorpdfstring{C$_2$H + HC${_3}$N $\rightarrow$ HC$_{5}$N + H}{reaction2} \label{subsec:overview-reaction2}}

This reaction is included in KIDA with two different estimated values ($\alpha = 1.06 \times 10^{-10}$; $\beta = -0.25$ for 10-280 K and  $\alpha = 1.3 \times 10^{-10}$; $\beta = 0$ for 50-200 K). In both cases, HC$_{5}$N + H is the sole reaction channel considered. In the last release 
of the UDfA \citep{Umist2022}, the reaction is included with the experimental rate coefficient obtained with the CRESU technique and that we are reporting in this manuscript for the first time. Again, HC$_{5}$N + H is the sole reaction channel considered.

To the best of our knowledge, there are no prior theoretical studies of the reactive PES. Therefore, we carried out QM calculations to derive the PES and estimate the rate coefficients and product BFs. 
The experimental results are reported in Sec.~\ref{subsubsec:experiments-results-reaction2} and, in more detail, in the Appendix \ref{subsec:append-exp-R2}, while the results from the theoretical calculations are reported in Sec.~\ref{subsubsec:theo-results-reaction2}.

\subsection{\label{subsec:overview-reaction3}
Reaction [3]: \texorpdfstring{CN + C$_{4}$H$_{2}$  $\rightarrow$ HC$_{5}$N + H}{reaction3}}

The reaction between cyano radicals (CN) and diacetylene (C$_{4}$H$_{2}$) has been first proposed by \cite{herbst1990gas} as a viable route for the formation of HC$_5$N. It has been the object of several theoretical \citep{fukuzawa1998neutral,tang2007dft,zhang2009crossed,huang2000crossed} and experimental studies \citep{seki1996reaction,zhang2009crossed,huang2000crossed}.

The rate coefficient of the CN + C$_{4}$H$_{2}$  reaction was measured experimentally by \cite{seki1996reaction} at 298 K. A large value of $4.2\times10^{-10}$ cm$^3$ s$^{-1}$ was derived. However, it is well established that the temperature dependence could significantly affect the value of the rate coefficient. Before our study, this information was missing.

 According to the PES calculations, confirmed by the results of a crossed molecular beam experiment, there is only one open channel (the one leading to HC$_{5}$N + H) at low temperatures (or collision energies) but there is more than one barrierless addition mechanism  \citep{fukuzawa1998neutral,tang2007dft,zhang2009crossed,huang2000crossed}.
 
Despite the presence of previous theoretical studies, with the aim to benchmark our method, we carried out electronic structure calculations and characterised the PES and the rate coefficient to be compared to the experimental one. 
The experimental results are described in Sec.~\ref{subsubsec:experiments-results-reaction3}
and, in more detail, in the Appendix \ref{subsec:append-exp-R3}, while the results from the theoretical calculations are reported in Sec.~\ref{subsubsec:theo-results-reaction3} and in the Appendix \ref{subsec:append-theo-R3}.

\subsection{Reactions [4]: \texorpdfstring{C$_{4}$H + HCN or HNC $\rightarrow$  HC$_{5}$N + H}{reaction4} \label{subsec:overview-reaction4}}

For reactions [4a] and [4b] there are no experimental/theoretical data available in the literature. They are included in the KIDA and UDfA databases but with different roles and estimated parameters. Only the reaction between the butadiynyl radical (C$_{4}$H) and HCN [4a] is included in the KIDA database as a possible formation route of HC$_5$N in planetary atmospheres, but not in interstellar objects. A very small rate coefficient is assumed by analogy with the value for the C$_2$H$_3$ + HCN reaction ($k(T)=4.5\times10^{-14}$ cm$^3$ s$^{-1}$). The UDfA database, instead, considers reaction [4b] with a large rate coefficient ($k(T)=2.5\times10^{-10}{\left(\frac{T}{300~\text{K}}\right)}^{-0.2}$ cm$^3$ s$^{-1}$), without any explanation on the assumed value.
Given the possible role of these two reactions in the formation of HC$_5$N and the lack of theoretical or experimental studies, we performed QM calculations to characterize the PES and estimate the rate coefficients. 
The results are reported in Sec.~\ref{subsubsec:theo-results-reaction4} and in the Appendix \ref{subsec:append-theo-R4}.

\subsection{Reaction [5]: \texorpdfstring{C${_6}$H + N $\rightarrow$ HC$_{5}$N + C}{reaction5} \label{subsec:overview-reaction5}}

The reactions between polyynyl radicals C$_6$H, C$_8$H, and C$_{10}$H with N atoms have been proposed by \cite{loison2013gas} to explain the formation of cyanopolyynes up to HC$_9$N. For the reaction C${_6}$H + N, two product channels have been suggested to be open: one leading to HC$_5$N + C ($\Delta H$ = $-$170 kJ mol$^{-1}$) and one leading to C$_6$N + H ($\Delta H$ = $-$227 kJ mol$^{-1}$). The global rate coefficients have been estimated using capture theory, while the product BFs have been guessed by considering the relative exothermicity of the two channels. 
Since \cite{loison2013gas} did not derive the entire reaction PES or provide statistical estimates of the product branching fractions, we have carried out dedicated calculations. 
We also note that the reaction is reported in the KIDA database, but not in the UdFA database.
The results are reported in Sec.~\ref{subsubsec:theo-results-reaction5} and in the Appendix \ref{subsec:append-theo-R5}.

\subsection{Reactions [6]: \texorpdfstring{C${_5}$H$_2$ + N $\rightarrow$ HC$_{5}$N + H}{reaction6} \label{subsec:overview-reaction6}}

The reaction between (C${_5}$H$_2$)  and nitrogen atoms has never been studied experimentally or theoretically. It is listed in the KIDA and UDfA databases with an estimated rate coefficient 
of $k(T)=1\times10^{-13}$ cm$^3$ s$^{-1}$, 
based on laboratory experiments involving similar systems. 
Our astrochemical model suggests that the contribution of this reaction to the formation of HC$_5$N in cold clouds is negligible. To significantly impact the models, an increase of several orders of magnitude in the value of  $k(T)$ would be required. Since such an increase is unrealistic, we have decided not to pursue further investigation of this reaction.

\subsection{Reactions [7]: \texorpdfstring{HC$_{5}$NH$^+$ + $e^-$ $\rightarrow$  HC$_{5}$N + H}{reaction7} \label{subsec:overview-reaction7}}

The dissociative electron-ion recombination of protonated cyanobutadiene (HC$_{5}$NH$^+$) is the final step in any ionic pathway that leads to the formation of HC$_{5}$N. As there are no experimental data available on the rate coefficients or BFs, 
the KIDA database reports a rate coefficient derived by 
considering similar systems. \citep{geppert2004dissociative,vigren2009dissociative}. The recommended rate coefficient and its dependence on the temperature is reported as $k(T)=2.0\times10^{-6}{\left(\frac{T}{300~\text{K}}\right)}^{-0.7}$ cm$^3$ s$^{-1}$. This suggested value, 
extrapolated from the dissociative recombination reaction of DC$_3$ND$^+$ and C$_2$H$_3$CNH$^+$,
is slightly larger than the one measured for HCNH$^+$ in line with the fact that dissociative
recombination reactions are faster when the complexity of the ion increases \citep{semaniak2001dissociative} and exhibits a significant temperature dependence, which is typical of dissociative recombination reactions of nitriles \citep{vigren2008dissociative,mclain2009flowing}.
Regarding the BFs, half of the products are assumed to preserve the carbon-nitrogen-chain (HC$_5$N or C$_5$N), while the other half 
involves breaking the carbon-nitrogen skeleton (HC$_3$N, HCN, C$_4$H, C$_2$H), as observed for other nitriles \citep{geppert2007formation}. The values of the BFs are provided in Tab.~\ref{tab:revised-network}. Given the reliability of these estimated values, we have decided not to revise this reaction and will utilize the database values in the updated network.     

\subsection{Reactions [8]: \texorpdfstring{H$_3$C$_{5}$N$^+$ + $e^-$ $\rightarrow$  HC$_{5}$N + H$_2$}{reaction8} \label{subsec:overview-reaction8}}

This dissociative electron-ion recombination reaction has been proposed by \cite{Herbst1989ApJS}. 
Since there are no theoretical or experimental studies available, a rate coefficient of $\sim2.0\times10^{-6}$ cm$^3$ s$^{-1}$ was assumed. 
A slightly negative temperature dependence was inferred by analogy with similar reactions involving smaller species
In terms of potential products, the authors only considered product channels that involve breaking one or two C--H or N--H bonds with equal probability,
leading to the formation of C$_5$N + H + H$_2$ or HC$_5$N + H$_2$. 
Since, in all cases, the contribution of this reaction is negligible for the formation of HC$_5$N, we decided not to revise it and to adopt the values reported in the databases.

\subsection{Literature search summary} \label{subsec:overview-summary}

We have reviewed 
nine reactions that may contribute to the formation of HC$_5$N, as documented in the literature and/or in the KIDA and UDfA astrochemical databases.
After a first run of our astrochemical model described in Sec.~\ref{sec:astro-modeling}, we verified that reactions [6] and [8] play a negligible role in the formation of HC$_5$N. In the case of reaction [7], we retained the value suggested in KIDA after a critical revision of the available data on similar systems. 

Instead, because of the incompleteness of the data and the potential lack of accuracy of the model outcomes, we have carried out:
\begin{itemize}
    \item [-] CRESU experiments to determine the global rate coefficients as a function of the temperature, down to 24 K, for the reactions [1], [2] and [3];
    \item [-] dedicated electronic structure calculations of the potential energy surfaces for the reactions [1], [2], [3], [4a], [4b] and [5]; rate coefficients and product BFs have also been calculated.
\end{itemize}

\section{Experimental and theoretical methods to derive the reaction rate coefficients and product branching fractions}\label{sec:methods}

\subsection{The CRESU technique}\label{subsec:cresu}

The CRESU technique was used to generate the low temperature collisional environment in which the kinetics of reactions [1]-[3] could be studied, and has been described in detail previously \citep{RN369,RN587,RN118}. A brief overview is given here. The CRESU core principle is the generation of a cold uniform supersonic flow, from which it obtains its name (Cinétique de Réaction en Ecoulement Supersonique Uniforme, or Reaction Kinetics in Uniform Supersonic Flow). A continuously fed gas reservoir is located in a low pressure, continuously pumped chamber. At the interface between the chamber and the reservoir, a Laval nozzle generates an isentropic expansion of a buffer (or carrier) gas. This expansion in an axisymmetric nozzle generates a uniform, low temperature, high-density gas flow. This dense expansion ensures that species are maintained in thermal equilibrium. While some pulsed variants of the CRESU have been developed in order to reduce gas consumption and pumping requirements \citep{RN138,RN601,RN600,RN1216,RN450}, here the flow is continuous. One major advantage is the very high quality of the flows generated (both in terms of available hydrodynamic time and variations of the temperature in the flow) at the expense of high gas consumption (typically 30-80 L/min of carrier gas). Adjustments to the chamber and reservoir pressure are required to correctly operate the system and generate the flow. Proper adjustment allows for flow durations in the range [200-2000] $\mu$s depending on the nozzle.
As the molecular radicals involved in reactions [1]-[3] are unstable, they must be generated in situ, and we have chosen to use pulsed laser photolysis (PLP) for this purpose. Besides the carrier gas, small amounts (less than 0.01\%) of a chemical precursor (for the radical) and co-reactant (less than 1\%) are introduced into the flow. Specific details of these precursors, as well as details of the detection schemes, are given in the Appendix \ref{sec:appendix-results} for each of the reactions under study.

\subsection{Theoretical Methods} \label{subsec:theory-methods}
\subsubsection{Electronic structure calculations} \label{subsubsec:theo-methods-pes}

The potential energy surfaces of the systems under investigation were characterized by optimizing the stationary points along the possible reaction pathways. The geometries of minima and saddle points were initially optimized using the hybrid M06-2X functional \citep{zhao2008m06} coupled with the correlation consistent valence polarized basis set aug-cc-pVTZ \citep{dunning1989gaussian}. The nature of each stationary point, identified through geometry optimization, was determined by conducting harmonic vibrational frequency calculations. For M06-2X calculations, Grimme’s D3 dispersion term was included during the geometry optimizations \citep{grimme2010consistent}. Additionally, all the saddle points have been characterized using Intrinsic Reaction Coordinate (IRC) calculations \citep{gonzalez1990reaction,gonzalez1989improved}.
To obtain more precise energy values for each stationary point, coupled-cluster single and double excitations augmented by a perturbative treatment of triple excitations (CCSD(T)) \citep{bartlett1981many,raghavachari1989fifth,olsen1996full} calculations were performed with the same basis set. The zero-point energy (ZPE) correction, computed using harmonic vibrational frequencies obtained at the M062X/aug-cc-pVTZ level of theory, was added to the CCSD(T) energies to correct them at 0 K. The accuracy of the calculated energies is expected to be within $\pm$5\kjm.
All calculations were carried out using the GAUSSIAN 09 \citep{g09} and GAUSSIAN 16 \citep{g16} packages. 
Furthermore, to verify the correctness of some selected stationary points that might be relevant for the kinetics, more accurate calculations were performed at the CCSD(T) level corrected with a Density Fitted (DF) MP2 extrapolation to the complete basis set (CBS) and with corrections for core electrons excitations. The CBS calculations were done using MOLPRO \citep{werner2020molpro}.

\subsubsection{Kinetics calculations} \label{subsubsec:theo-methods-kinetics}

Kinetics calculations were performed by using a homemade code \citep{balucani2009combined,balucani2010formation} based on capture and Rice–Ramsperger–Kassel–Marcus (RRKM) theories, as detailed in prior works \citep{vazart2015cyanomethanimine,skouteris2018genealogical,skouteris2019interstellar,giani2023revised}.
The initial bimolecular rate coefficient, leading from reactants to the first intermediate, was determined through capture theory, which assumes for neutral-neutral reactions an entrance potential described by the equation:
\begin{equation}
    V(R)= -\frac{C_6}{R^6}
	\label{eq:capture}
\end{equation}
where $R$ is the distance between the two species and $C_6$ is the interaction coefficient for a dipole-dipole pair, determined by fitting long-range \textit{ab initio} data. The rate of the back dissociation is calculated from the capture rate coefficient and the ratio between the densities of states of the reactants and the initial intermediate.
%
%

The microcanonical rate coefficient of each unimolecular step is calculated using the formula:
\begin{equation}
    k_{E}= \frac{N(E)}{h\rho(E)}
\end{equation}
where $N(E)$ is the sum of states of the transition state at energy $E$, $\rho(E)$ is the reactant density of states at energy $E$ and $h$ is Planck's constant. Having the rate coefficients for each unimolecular step, it is possible to solve the master equation and derive the energy-dependent rate coefficients for the formation of each product. Finally, Boltzmann averaging is performed to obtain canonical (temperature-dependent) rate coefficients.

The rate coefficients were then fitted to the modified Arrhenius form used in the astrochemical databases and models:
\begin{equation} \label{eq:Arrhenius}
   k(T)=\alpha \left(\frac{T}{300~\text{K}}\right)^\beta ~exp\left(-\frac{\gamma}{T}\right)
\end{equation}


\section{Main results from the CRESU experiments} \label{sec:experiments-results}

\subsection{Main results for reaction [1]: \texorpdfstring{C${_3}$N + C$_{2}$H$_{2}$  $\rightarrow$  HC$_{5}$N + H}{reaction1} \label{subsubsec:experiments-results-reaction1}}

The rate coefficient for this reaction has been measured between 24 and 294 K. It is fast in the explored temperature range and exhibits a
negative temperature dependence. The resulting global expression of the rate coefficient as a function of the temperature is

\begin{equation}
    k_{\text{C$_3$N + C${_2}$H${_2}$}}(T)= 3.091\times10^{-10}{\left(\frac{T}{300~\text{K}}\right)}^{-0.577}   
    \exp{\left(\frac{-33.64~\text{K}}{T}\right)}
    ~~~ \text{cm}^3 \text{s}^{-1}
\end{equation}

All the relevant details on the experimental conditions, detection scheme and method to produce the C$_3$N radicals are given in Appendix \ref{subsec:append-exp-R1}.

\subsection{Main results for reaction [2]: \texorpdfstring{C$_2$H + HC${_3}$N $\rightarrow$ HC$_{5}$N + H}{reaction2} \label{subsubsec:experiments-results-reaction2}}
The rate coefficient for this reaction has been measured between 24 and 297~K. It is fast in the explored temperature range, with a strong negative temperature dependence:

\begin{equation}
    k_{\text{C$_2$H + HC${_3}$N}}(T)= 3.91\times10^{-11}{\left(\frac{T}{300~\text{K}}\right)}^{-1.04}   ~~~ \text{cm}^3 \text{s}^{-1}
\end{equation}

In general, this is an indication of a barrierless (or at least only with submerged barriers) exothermic reaction. 


All the relevant details on the experimental conditions, detection scheme, and method to produce the C$_2$H radicals are given in Appendix \ref{subsec:append-exp-R2}.

\subsection{Main results for reaction [3]: \texorpdfstring{CN + C$_{4}$H$_{2}$ $\rightarrow$ HC$_{5}$N + H}{reaction3} \label{subsubsec:experiments-results-reaction3}}

The rate coefficient has been measured between 24 and 300~K. The resulting global $k$ expression as a function of the temperature is

\begin{equation}
    k_{\text{CN + C${_4}$H$_2$}}(T)= 4.06\times10^{-10}{\left(\frac{T}{300~\text{K}}\right)}^{-0.24} \exp{\left(\frac{-11.5~\text{K}}{T}\right)} ~~~ \text{cm}^3 \text{s}^{-1}
\end{equation}

All the relevant details on the experimental conditions, detection scheme, and method to produce the CN radicals are given in Appendix \ref{subsec:append-exp-R3}.


\section{Main results from the theoretical calculations} \label{sec:theo-results}

In this section, we report a summary of the theoretical results on the PESs and reaction rate coefficients. Some details will be provided only for reaction [2] taken as an example. The results for the other five reactions ([1], [3], [4a], [4b] and [5]) are detailed in Sec.~\ref{sec:appendix-results-theo} of the Appendix.

\subsection{Reaction [1]: \texorpdfstring{C${_3}$N + C$_{2}$H$_{2}$  $\rightarrow$  HC$_{5}$N + H}{reaction1} \label{subsubsec:theo-results-reaction1}}

Our calculations (for the details, see Sec.~\ref{subsec:append-theo-R1}) demonstrate that HC$_{5}$N is the main product of the reaction, with a BF $\sim$1.0 in the investigated temperature range (10--400~K). Our calculated rate coefficient is consistent with the experimental one, being lower by a factor of 1.2 at 24~K and a factor of 1.6 at 100~K (see Tab.~\ref{tab:comp-exp-theo}). 
Moreover, at 10 K the calculated rate coefficient ($k_{\text{this work}}$= 1.4$\times$10$^{-10}$ cm$^3$ s$^{-1}$) is consistent with the KIDA value ($k_{\text{KIDA}}$= 1.3$\times$10$^{-10}$ cm$^3$ s$^{-1}$).

\subsection{Reaction [2]: \texorpdfstring{C$_2$H + HC${_3}$N $\rightarrow$ HC$_{5}$N + H}{reaction2} \label{subsubsec:theo-results-reaction2}}

\begin{figure*}
 	\includegraphics[width=9cm]{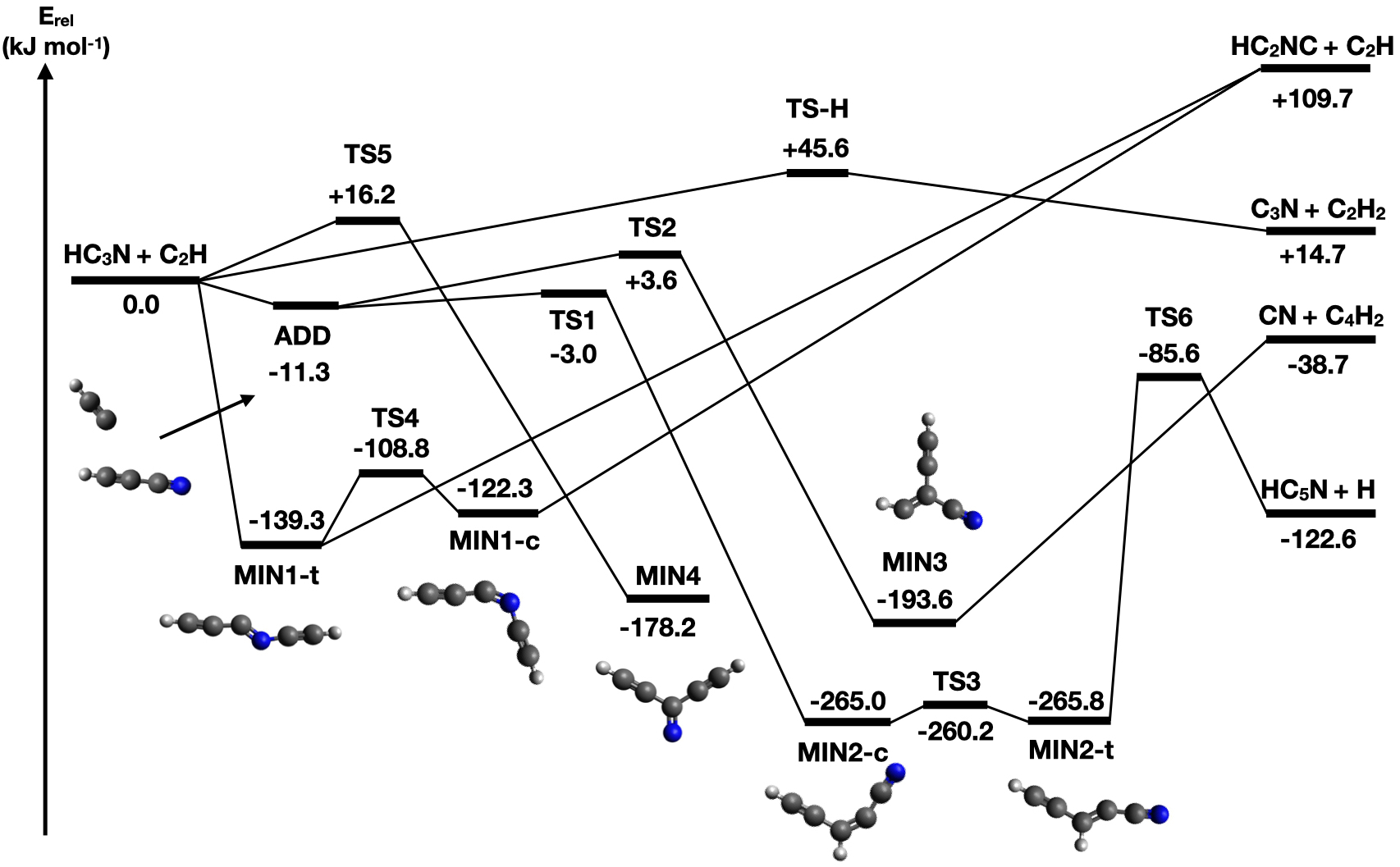}
   	\includegraphics[width=8cm]{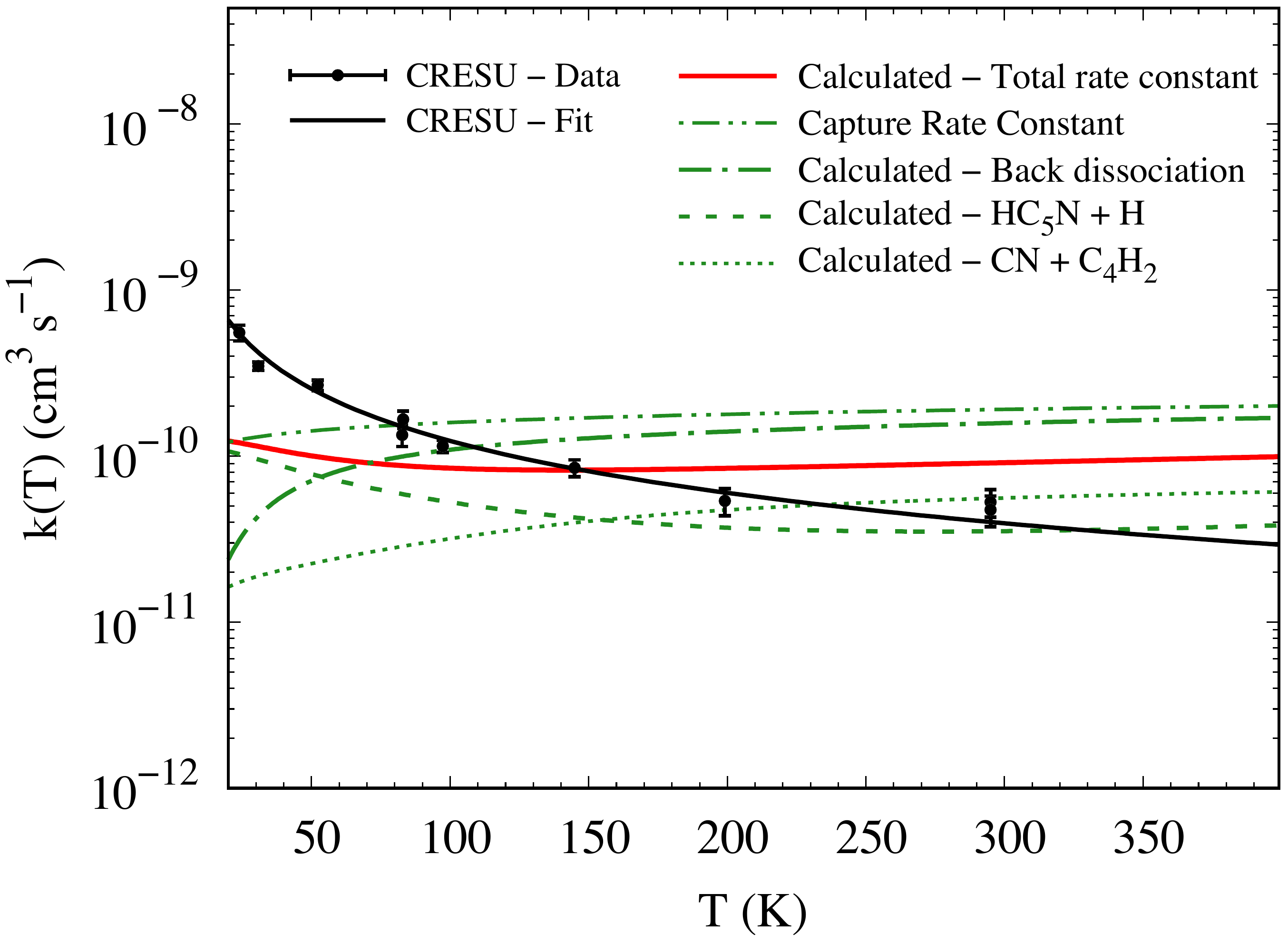}
    \caption{\textit{Left panel}: PES of the C$_2$H + HC${_3}$N reaction ([2] of Tab.~\ref{tab:revised-network}).
    Electronic structure calculations of the stationary points were performed at the CCSD(T)/aug-cc-pVTZ//M06-2X/aug-cc-pVTZ level of theory. \textit{Right panel}: rate coefficients for the reaction [2] as a function of temperature (10-400 K). 
    The black dots are the CRESU experimental data, while the black line represents their fit. 
The red solid line represents the calculated reaction rate coefficient (see Sec.~\ref{subsubsec:theo-results-reaction2}). 
    The green lines represent the contribution of the different channels to the total rate coefficient: capture coefficient (dashed-dot-dot), HC$_5$N + H (dashed), CN + C$_4$H$_2$ (dotted) and back-dissociation (dotted-dashed).}
    \label{fig:pes-rate-reac-2}
\end{figure*}

\subsubsection{PES of reaction [2]}

The calculated PES for the reaction [2] is shown in Fig.~\ref{fig:pes-rate-reac-2}. Two different initial steps are possible: H-abstraction with the direct formation of C$_2$H$_2$ + C${_3}$N or addition of C$_2$H on different sites of HC$_3$N. The H-abstraction channel is endothermic by 14.7~\kjm and presents a barrier of 45.6~\kjm (TS-H). Therefore, this reaction pathway cannot be competitive under the conditions of interstellar objects. 
Seven minima (ADD, MIN1-c, MIN1-t, MIN2-c, MIN2-t, MIN3 and MIN4) and seven transition states (TS1, TS2, TS3, TS4, TS5, TS6 and TS7) were identified along the addition pathway. 
The addition of C$_2$H on the C$\equiv$C bond of HC$_3$N starts with the barrierless formation of a van der Waals (vdW) adduct (ADD), with a C-C distance equal to 2.81 \AA  \ (the stabilization of the adduct is 11.3~\kjm with respect to the reactants energy asymptote at the CCSD(T)/aug-cc-pVTZ level of theory). 
The adduct easily converts into the covalently bound MIN2-c ($-$265.8~\kjm) by overcoming a barrier (TS1) which is located slightly above the adduct but below the reactants (-3.0~\kjm) at the level of theory employed. MIN2-c can isomerise from the \textit{cis} to the \textit{trans} isomer (MIN2-t) by overcoming a small barrier of 4.8~\kjm (TS3). The fission of the middle C-H bond in MIN2-t leads to the formation of HC$_5$N + H, with a global enthalpy change of $-$122.6~\kjm (a barrier of 180.2~\kjm (TS6) needs to be overcome). 
The vdW adduct can also rearrange into MIN3 through TS2, which is located at +3.6~\kjm above the energy of the reactants. The presence of TS1 and TS2 is associated with a change of hybridization (from sp to sp$^2$) of the carbon atom to which C$_2$H adds.
A C-C bond-breaking mechanism in MIN3 leads to the formation of C$_4$H$_2$ + CN (the enthalpy change for this channel is $-$38.7~\kjm) in a barrierless process (confirmed by scans performed along the reaction coordinate). 
The addition of C$_2$H can also occur on the cyano group of HC$_3$N: the addition on the carbon atom through TS5 (+16.2~\kjm) leads to the formation of MIN4 ($-$178.2~\kjm), which presents a symmetric structure.
MIN4 can dissociate back to the reactants, or it can lose an H atom or a CH group and form HC$_3$(N)C$_2$ + H ($\Delta H$ = +368.8~\kjm) and HC$_3$(N)C + CH products ($\Delta H$ = +274.5~\kjm), respectively. However, given their large endothermicity, these two channels are not open under the conditions of interstellar objects. 
Finally, the C$_2$H addition on the nitrogen atom leads to the barrierless formation of MIN1-t ($-$139.3\kjm), which easily isomerises into MIN1-c through TS4 (barrier of 30.5~\kjm). Both MIN1-t and MIN1-c can evolve further only by a C-C bond breaking mechanism that leads to the endothermic formation of HC${_2}$NC + C$_2$H products, ($\Delta H$ = +109.7~\kjm). Therefore, also this channel cannot contribute.\\
\indent
 Considering the expected accuracy of our calculations, we cannot exclude that both TS1 and TS2 
 are either slightly above or below the reactants energy asymptote. 
 


\subsubsection{Kinetics of reaction [2]}

The calculated rate coefficients for the reaction [2] are shown in Fig.~\ref{fig:pes-rate-reac-2}. 
According to our electronic structure calculations, the reaction is expected to occur through the initial formation of ADD, followed by the isomerisation to MIN2-c.
The back-dissociation dominates the reaction since the first complex formed (ADD) is only 11.3~\kjm more stable than the reactants and the successive step, the isomerisation to MIN2-c through TS1, has to overcome a barrier at an energy comparable to that of the reactants (-3.0~\kjm). At low temperatures (10-150 K), the main products are HC$_5$N + H, while the formation of CN + C$_4$H$_2$ is hindered by the presence of TS2 at 3.6\kjm above the energy of the reactants. However,  when the temperature approaches 150 K, the CN + C$_4$H$_2$ channel becomes as important as the  HC$_5$N + H channel. 
The H-abstraction channel can be neglected since it is characterized by a significant entrance barrier (45.6~\kjm). The formation of MIN4 is also neglected since it can only rearrange back to the reactants or form products associated with channels with large endothermicity. 
The total rate coefficient shows a negative temperature dependence at low \textit{T} (<150 K) because of the increasing contribution of the back dissociation. The $\alpha$, $\beta$ and $\gamma$ coefficients derived from the fitting of the rate coefficients of formation of HC$_5$N + H and CN + C$_4$H$_2$ are reported in Tab.~\ref{tab:revised-network}, while a comparison of the calculated total rate coefficient and the experimental one is reported in Tab.~\ref{tab:comp-exp-theo}. These results are consistent with the experimental value around 150 K, but differ at lower and higher temperatures. In particular, at 24~K the calculated rate coefficient is a factor 5.6 lower than the experimental one and at 100~K it is a factor 1.4 lower. The comparison between experimental and theoretical values suggests that the vdW adduct resides in a deeper well with respect to the one we have calculated and/or the energy barrier associated with TS1 is overestimated. 
A slightly larger value of the TS2 energy could, instead, reconcile theoretical predictions and experimental values at high temperatures. Calculations at an increased level of theory could address these points, but they are out of the scope of the present work. Overall the agreement between theoretical and experimental values is acceptable. 
Furthermore, the calculated rate coefficient at 10 K ($k_{\text{this work}}$= 1.5$\times$10$^{-10}$ cm$^3$ s$^{-1}$) is consistent with the KIDA value ($k_{\text{KIDA}}$= 2.5$\times$10$^{-10}$ cm$^3$ s$^{-1}$).

\subsection{Reaction [3]: \texorpdfstring{CN + C$_{4}$H$_{2}$ $\rightarrow$ HC$_{5}$N + H}{reaction3} \label{subsubsec:theo-results-reaction3}}

The QM calculations are in line with previous work (for the details see Sec.~\ref{subsec:append-theo-R3}) and confirm that the only exothermic channel is the one leading to the formation of HC$_{5}$N. Therefore, at low temperatures (<400 K) HC$_{5}$N is the main product of the reaction. The calculated rate coefficients are consistent with the experimental values, being a factor 2.1 lower at 24~K and 1.7 at 100~K (see Tab.~\ref{tab:comp-exp-theo}). 
Moreover, at 10 K the calculated rate coefficient ($k_{\text{this work}}$= 1.8$\times$10$^{-10}$ cm$^3$ s$^{-1}$) is consistent with the KIDA value ($k_{\text{KIDA}}$= 2.4$\times$10$^{-10}$ cm$^3$ s$^{-1}$).

\subsection{Reactions [4a] and [4b]: \texorpdfstring{C$_{4}$H + HCN or HNC $\rightarrow$  HC$_{5}$N + H}{reaction4} \label{subsubsec:theo-results-reaction4}}

According to the calculations (see Sec.~\ref{subsec:append-theo-R4}), HC$_{5}$N + H is the main channel for both reactions [4a] and [4b]. However, the rate coefficient for reaction [4b] is larger than that for reaction [4a]. This is due to the presence of vdW adducts and transition states with an energy comparable to the energy of the reactants in the case of reaction [4a], which causes significant back-dissociation.  
The formation of HC$_{4}$NC is possible in both reactions but its yield is negligible in the investigated temperature range (10-400 K). 
For reaction [4a], the rate coefficient calculated at 10 K ($k_{\text{this work}}$= 5.8$\times$10$^{-11}$ cm$^3$ s$^{-1}$) is three orders of magnitude larger than the KIDA value ($k_{\text{KIDA}}$= 4.5$\times$10$^{-14}$ cm$^3$ s$^{-1}$). For reaction [4b], the calculated rate coefficient at 10 K is ($k_{\text{this work}}$= 1.5$\times$10$^{-10}$ cm$^3$ s$^{-1}$) is consistent with the UDfA value ($k_{\text{UDfA}}$= 5$\times$10$^{-10}$ cm$^3$ s$^{-1}$).

\subsection{Reaction [5]: \texorpdfstring{C${_6}$H + N $\rightarrow$ HC$_{5}$N + C}{reaction5} \label{subsubsec:theo-results-reaction5}}

According to our calculations (see Sec.~\ref{subsec:append-theo-R5}) the channel leading to HC$_{5}$N + C is by far the dominant channel with a BF of $\sim$1.0. The yields of the C$_6$N + H and C$_5$H + CN channels are negligible in the investigated temperature range (10-400 K).
The total rate coefficient ($k^{10~\text{K}}_{\text{tot, this work}}$= 1.7$\times$10$^{-10}$ cm$^3$ s$^{-1}$) is consistent with the one calculated by \cite{loison2014interstellar} ($k^{10~\text{K}}_{\text{tot, KIDA}}$= 9$\times$10$^{-11}$ cm$^3$ s$^{-1}$), but the BFs (estimated on the basis of the enthalpy change of the possible channels by \cite{loison2014interstellar}) are very different.

\section{Astronomical observations} \label{sec:observations}

In order to add constraints to the possible formation routes of HC$_5$N, we carried out new and retrieved old observations towards two regions representative of cold ($\sim$ 10 K) and warm ($\sim$ 90 K) environments: the cold core L1544 and the warm molecular shock L1157-B1, respectively.

L1544, situated in the Taurus molecular cloud complex at a distance of 170 pc \citep[e.g.][]{galli2019}, is regarded as the prototype of prestellar cores, on the brink of gravitational collapse \citep[e.g.,][]{caselli2012}. 
Its core exhibits a high density ($\sim$10$^{6}$ cm$^{-3}$) and a very low temperature ($\sim$7 K), leading to the unique chemistry associated with cold, CO-depleted gas, including significant species deuteration \citep[e.g.][]{caselli1999,crapsi2007,ceccarelli2007,caselli2022}. In contrast, the outer layers undergo diverse chemical processes that result in the formation of iCOMs, small and complex carbon chains \citep[e.g.][]{bizzocchi2014, vastel2014, vastel2016,Jimenez-Serra2016-L1544, punanova2018,Ceccarelli2023-PP7, spezzano2017observed, urso2019c2o, bianchi2023cyanopolyyne}. 
In addition to TMC-1, L1544 is one of the few cold cores where long cyanopolyynes up to HC$_9$N have been detected \citep{bianchi2023cyanopolyyne}.

L1157-mm is a low-mass Class 0 protostar located at a distance of 352 pc \citep{zucker2019GAIA}, and driving a powerful molecular outflow. 
Its blue-shifted lobe contains three chemically rich clumps: B0, B1, and B2, situated at the apex and along the walls of cavities created by a precessing jet \citep[e.g.][]{gueth1996, bachiller2001, podio2017silicon}. 
The B1 clump has been the focus of several molecular line studies, and turns out to be an ideal laboratory for astrochemical studies due to its molecular richness linked to the shocks in the region.
Because of them, several components of grain mantles are release into the gas-phase, via sputtering and shattering of dust grains, where they undergo subsequent reactions which enrich the chemical complexity \citep[e.g.][]{arce2008complex, lefloch2017, codella2017seeds, codella2020seeds}.
Relevant to the context of this article, cvyanopolyynes up to HC$_5$N have been detected in the L1157-B1 \citep{Bachiller1997-HC3NinL1157B1, 2018Mendoza}.

\subsection{Observations and results} \label{subsec:obs-results}

\subsubsection{L1544, a cold (10 K) prestellar core}

\cite{bianchi2023cyanopolyyne} detected several cyanopolyynes, from HC$_3$N to HC$_9$N, towards the prestellar core L1544, via observations obtained with the 100 m Robert C. Byrd Green Bank Telescope (GBT) in X band (8.0-11.6 GHz) and Ku band (13.5-14.4~GHz).
In order to constrain further the formation routes of HC$_5$N, we reconsidered the data published in \citet{bianchi2023cyanopolyyne} and searched for the emission of the $^{13}$C isotopologues. We refer to \citet{bianchi2023cyanopolyyne} for further details regarding the observations.

\begin{figure}
	\includegraphics[width=7.5cm]{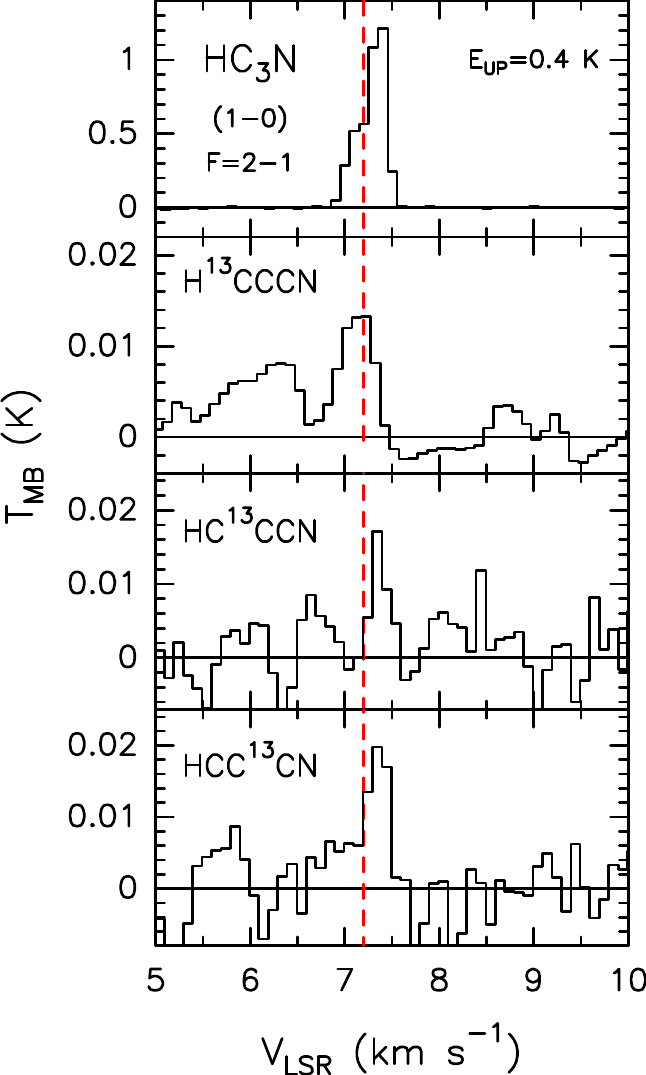}
    \caption{Lines from the different HC$_{\rm 3}$N isotopologues observed towards L1544 with GBT. 
    The vertical dashed lines mark the ambient local system rest (LSR) velocity (+7.2 km s$^{-1}$; \citealt{Tafalla1998}). The transition and its upper-level energy is reported on upper panel.}
    \label{fig:spectra-L1544}
\end{figure}

Unfortunately, we did not detect any $^{13}$C isotopologue of HC$_5$N. The brightest HC$_5$N transition is the (4-3) and the same transitions of H$^{13}$CCCCCN is not detected at 3$\sigma$, resulting in a lower limit of $^{12}$C/$^{13}$C $>$ 75, which is not significant enough to give useful constraints.
We also checked the lines from the HC$_3$N isotopologues, in order to have information on the possible signature of HC$_3$N as mother species of HC$_5$N.
We detected one transition from each of the HC$_3$N isotopologues: H$^{13}$CCCN, HC$^{13}$CCN, and HCC$^{13}$CN (see Tab.~\ref{Tab:lines-L1544}). 
The spectra are smoothed to a common spectral resolution of 0.1~km~s$^{-1}$. 
The detected lines are close in frequency and the rms noise varies between 3 and 4~mK in the considered spectral channel. 
The line corresponding to the main isotopologue is very bright with a main-beam temperature of 1.2~K and it is detected with a signal-to-noise ratio larger than 300.
The H$^{13}$CCCN, HC$^{13}$CCN, and HCC$^{13}$CN lines are detected with a signal-to-noise ratio of 4, 4 and 5, respectively.
The observed spectra are reported in Fig.~\ref{fig:spectra-L1544}.
The integrated intensity reported in Tab.~\ref{Tab:lines-L1544} is obtained integrating in the velocity range +6.9~km~s$^{-1}$ --7.55~km~s$^{-1}$. 
The range is derived from the bright HC$_{3}$N transition and applied to the other isotopologues.

\begin{table*}
	\begin{center}
 \renewcommand{\arraystretch}{1.2}
 \caption{List of transitions and line properties (in T$_{\rm MB}$ scale) of the HC$_{\rm 3}$N isotopologues emission. 
 The columns report the transition and their frequency (GHz), the upper level energy E$_{\rm up}$ (K), the line strength $S\mu^2$ (D$^2$), the line peak temperature (mK), the rms (mK), and the velocity integrated line intensity I$_{\rm int}$ (mK km s$^{-1}$). 
 The last column reports the ratio between the line peak temperature of each line with respect to that of the main isotopologue HC$_{\rm 3}$N.
\label{Tab:lines-L1544}}
\begin{tabular}{lccccccc}
\hline
Transition & $\nu$$^{\rm a}$ & E$_{\rm up}$$^{\rm a}$ & $S\mu^2$$^{\rm a}$ & T$_{\rm peak}$ & rms & I$_{\rm int}$$^{\rm b}$ & Ratio \\
 & (GHz) & (K) & (D$^2$) &  (mK) & (mK) & (mK km s$^{-1}$) & \\
\hline
HC$_{3}$N 1--0, F= 2--1 & 9.0983 & 0.44 & 7.7 & 1211 & 4 & 393 (1) & -- \\
\hline
H$^{13}$CCCN 1--0, F= 2--1 & 8.8171 & 0.42 & 7.7 & 13 & 3 & 5.2 (0.8) & 93 (21) \\
\hline
HC$^{13}$CCN 1--0, F= 2--1 & 9.0597 & 0.43 & 7.7 & 17 & 4 & 3.5 (1.13) & 71 (17) \\
\hline
HCC$^{13}$CN 1--0, F= 2--1 & 9.0606 & 0.43 & 7.7 & 20 & 4 & 6.8 (1.0) & 61 (12) \\
\hline
\end{tabular}\\
	\end{center}
$^{\rm a}$ Frequencies and spectroscopic parameters have been provided by \cite{1977Creswell} and retrieved from the Cologne Database for Molecular Spectroscopy \cite{Muller2001, Muller2005}.\\
$^{\rm b}$ Errors on the integrated intensity do not include 20$\%$ of calibration.\\     

\end{table*}

\subsection{Column densities and isotopologue ratios} \label{subsec:obs-ratios}

\subsubsection{L1544, a cold (10 K) prestellar core}
\citet{bianchi2023cyanopolyyne} measured a column density of HC$_5$N equal to (3--20) $\times$ $10^{13}$~cm$^{-2}$ and, based on the resolved line profiles, they attributed the cyanopolyynes emission with the southern region of the L1544 core.
Unfortunately, they could not estimate the H$_2$ column density of the HC$_5$N emitting region and, therefore, its abundance.
Nonetheless, the presence itself of HC$_5$N indicates that the region is mostly shielded by the UV photons, which gives a lower limit to N(H$_2$) $\geq 10^{22}$~cm$^{-2}$ and, consequently, HC$_5$N/H$_2$ $\leq 2\times10^{-8}$.
On the other hand, the H$_2$ column density cannot be larger than the L1544 column density at center, which is $\sim30$ times larger.
Therefore, the HC$_5$N abundance is  $1\times10^{-10} \leq $ HC$_5$N/H$_2$ $\leq 2\times10^{-8}$.

Finally, the abundance ratios, reported in Tab.~\ref{Tab:lines-L1544}, are derived from the ratios between the line peak temperatures.
They are HC$_{\rm 3}$N/H$^{13}$CCCN = $93\pm21$, HC$_{\rm 3}$N/HC$^{13}$CCN = $71\pm17$, and HC$_{\rm 3}$N/HCC$^{13}$CN = $61\pm12$. 


\subsubsection{L1157-B1, a warm (90 K) shocked region}

\cite{2018Mendoza} detected  HC$_3$N and HC$_5$N and the HC$_3$N $^{13}$C isotopologues towards the L1157-B1 region, from observations between 72 and 272~GHz obtained with the IRAM 30-m telescope.
Interestingly, the three HC$_3$N isotopologues show similar flux ratios within the errors (52$\pm$15, 63$\pm$20, and 81$\pm$10, for H$^{13}$CCCN, HC$^{13}$CCN, and HCC$^{13}$CN, respectively), indicating that there is not a differential $^{13}$C fractionation between the three HC$_3$N isotopologues, as in the case of L1544 (see above).
 
\cite{2018Mendoza} carried out a rotation diagram analysis of the HC$_5$N emission (that arises from the so-called \textit{g2} component of L1157-B1 corresponding to the outflow cavity walls \citep{lefloch2012ApJ}. They derived a HC$_5$N abundance with respect to H$_2$ equal to (0.8 -- 1.6) $\times 10^{-9}$. Finally, the relative abundance ratio HC$_3$N/HC$_5$N is equal to 3.
To the best of our knowledge, there are no other detections of HC$_5$N clearly associated with low-mass protostellar shocks.




\section{The revised gas-phase reaction network} \label{sec:revised-network}

Table \ref{tab:revised-network} summarizes the reactions, products, and rate coefficients of the new revised network for the formation of cyanobutadiyne.
The rate coefficients are given in terms of the parameters $\alpha$, $\beta$ and $\gamma$, following the Eq.~\ref{eq:Arrhenius}, which is commonly used in the astrochemical databases and models. 
It should be noted that $\alpha$, $\beta$ and $\gamma$ are the results of the fits of the rate coefficients as a function of the temperature, and the $\gamma$ parameter is sometimes different from zero to make the analytical fit better. However, in those cases, $\gamma$ must not be considered as an activation barrier.

The new network consists of nine reactions, forming HC$_5$N or one of the sets of products of the competing channels.
These are the main changes with respect to the previous assumptions:
\begin{itemize}
    \item [-] Reactions [1], [2] and [3] have been revised and the theoretical results have been compared to the experimental ones. We verified the impact of the different rate coefficients on the modelled abundance of HC$_5$N by using only the experimental results or only the theoretical results. The results obtained using the two networks are discussed in Sec.~\ref{subsubsec:comp-exp-theo}.
    \item [-] Reactions [4a] and [4b] have been revised and included in the network since HC$_{5}$N is the main product of both the reactions. The channel of formation of HC$_{4}$NC is negligible at the temperatures considered in this work (10-400 K). Note that reaction [4a] was present in KIDA for atmospheric conditions and not in UDfA, while reaction [4b] was present in UDfA and not in KIDA.
    \item [-] Reaction [6] has not been revised, since it is likely negligible in the formation of cyanobutadiyne.
    \item [-] Reactions [7] and [8] have not been revised in this work and the database values are used. Their revision is left to a future work.
\end{itemize}
%

\section{Astrochemical modeling} \label{sec:astro-modeling}

We ran astrochemical models to assess the role of the gas-phase reactions for the formation of HC$_5$N and to check whether the revised network (described in Sec.~\ref{sec:revised-network} and Tab.~\ref{tab:revised-network}) can reproduce better the observed abundances. We used the chemical code MyNahoon \citep{wakelam2005estimation,wakelam2010sensitivity}, which computes the evolution of the abundances of gaseous species as a function of time for a given temperature and density. 
The model does not consider reactions occurring on the grain surfaces, except for the formation of H$_2$.
We adopted the GRETOBAPE gas-phase reaction network \citep{tinacci2023-gretobape}, updated with the reactions of formation of HC$_5$N. 
The reactions [1]-[3] have been studied both theoretically and experimentally, and the results differ somewhat at very low temperatures.

%

%
\renewcommand{\arraystretch}{1.8}
\begin{landscape}
\begin{table}
    \begin{center}
    \caption{Revised network for the formation of HC$_5$N in the gas phase. For reactions [1]-[3], we only report the theoretical rate coefficients in the revised network columns.
    For each reaction, the rate coefficients are given in terms of $\alpha$ (in cm$^3$ s$^{-1}$); column 2), $\beta$ (column 3) and $\gamma$ (in K; column 4), following the formalism of Eq.~\ref{eq:Arrhenius}.
    The rate coefficients are valid for the range of temperature reported in column 5.
    Note that the number of each reaction refers to those used in the subsections of Sec.~\ref{sec:overview-gas-routes}.
    For comparison, columns 6 to 11 report the values listed in the KIDA \citep{wakelam2024} and UDfA \citep{Umist2022} databases, respectively.}
    \label{tab:revised-network}
	\begin{tabular}{ccrcl|ccccc||ccc|ccc}
		\hline
 \multicolumn{10}{c||}{Revised network} & \multicolumn{3}{c|}{KIDA} & \multicolumn{3}{c}{UDfA} \\
\multicolumn{5}{c}{Reaction} & $\alpha$ & $\beta$ & $\gamma$ & T [K] & BF (10K) & $\alpha$ & $\beta$ & $\gamma$ & $\alpha$ & $\beta$ & $\gamma$ \\
	\hline
		\hline
  \multicolumn{15}{l}{\textit{(i) Routes to HC$_5$N}}\\
1  & & C$_{3}$N + C$_2$H$_{2}$      & $\rightarrow$ & HC$_{5}$N + H      &  3.27$\times10^{-10}$  & 0.24   & -0.43 &   10-298  & 100\% &  1.3$\times10^{-10}$  &   0.0  & 0.0  &  3.091$\times10^{-10}$ & -0.577 & 33.64 \\
   & &                              & $\rightarrow$ & HC$_{3}$N + C$_2$H &           -            &    -   &   -   &   10-298  &   -   &  2.19$\times10^{-10}$ & 0.0    & 0.0  & - & - & - \\
2  & & HC$_{3}$N + C$_2$H           & $\rightarrow$ & HC$_{5}$N + H      &  6.56$\times10^{-11}$  & -0.26  & 2.33  &   10-298  & 100\% &  1.06$\times10^{-10}$ & -0.25  & 0.0  &  3.91$\times10^{-10}$  & -1.04  & 0.0   \\
3  & & C$_4$H$_{2}$ + CN            & $\rightarrow$ & HC$_{5}$N + H      &  3.58$\times10^{-10}$  & 0.2    & 0.0   &   10-298  & 100\% &  2.72$\times10^{-10}$ &  -0.52 & 19.0 & 4.06$\times10^{-10}$  & -0.24  & 11.5  \\
4a &*& C${_4}$H + HCN               & $\rightarrow$ & HC$_{5}$N + H      &  7.72$\times10^{-12}$  & -0.62  & 0.94  &   10-150  & 100\% &  4.5$\times10^{-14}$  &  0.0   & 0.0  & - & - & - \\ 
   & &                              & $\rightarrow$ & HC$_{5}$N + H      &  5.08$\times10^{-12}$  & 2.29   &-363.0 &  150-400  & 100\% &  4.5$\times10^{-14}$  &  0.0   & 0.0  & - & - & - \\ 
4b & & C${_4}$H + HNC               & $\rightarrow$ & HC$_{5}$N + H      &  2.99$\times10^{-10}$  &  0.21  &  0.0  &   10-300  &       &  -                    &  -     & -    & 2.5$\times10^{-10}$ & -0.2 & 0.0 \\
5  & & C${_6}$H + N                 & $\rightarrow$ & HC$_{5}$N + C      &  3.95$\times10^{-10}$  &   0.24 &  0.0  &   10-280  & 99.2\%& 6$\times10^{-11}$     &  0.17  & 0.0  & - & - & - \\
   & &                              & $\rightarrow$ & C$_{6}$N + H       &  3.16$\times10^{-13}$  &   0.19 &  0.0  &   10-280  & 0.7\% & 1$\times10^{-10}$     &  0.17  & 0.0  & - & - & - \\
   & &                              & $\rightarrow$ & C$_{5}$H + CN      &  4.06$\times10^{-12}$  & 0.55   & 0.0   &   10-280  &0.01\% &     -                 &  -     &  -   & - & - & - \\
6  & & C${_5}$H$_2$ + N             & $\rightarrow$ & HC$_{5}$N + H      &  1$\times10^{-13}$     &   0.0  &  0.0  &   10-280  & 100\% & 1$\times10^{-13}$     & 0.0    & 0.0  & 1$\times10^{-13}$  & 0.0  & 0.0 \\
7  & & HC$_{5}$NH${^+}$ + $e^-$     & $\rightarrow$ & HC$_{5}$N + H      & 9.2$\times10^{-7}$     &  -0.7  &  0.0  &   10-300  & 46\%  & 9.2$\times10^{-7}$    &  -0.7  & 0.0  & 9.2$\times10^{-7}$    &  -0.7  & 0.0 \\
   & &                              & $\rightarrow$ & C$_{4}$H + HCN     &  4.4$\times10^{-7}$    &  -0.7  &  0.0  &   10-300  & 22\%  & 4.4$\times10^{-7}$    &  -0.7  & 0.0  & 4.4$\times10^{-7}$    &  -0.7  & 0.0 \\
   & &                              & $\rightarrow$ & C$_{4}$H + HNC     &  4.4$\times10^{-7}$    &  -0.7  &  0.0  &   10-300  & 22\%  & 4.4$\times10^{-7}$    &  -0.7  & 0.0  & 4.4$\times10^{-7}$    &  -0.7  & 0.0 \\
   & &                              & $\rightarrow$ & C$_{2}$H + HC$_3$N &  1.2$\times10^{-7}$    &  -0.7  &  0.0  &   10-300  &  6\%  & 1.2$\times10^{-7}$    &  -0.7  & 0.0  & 1.2$\times10^{-7}$    &  -0.7  & 0.0  \\
   & &                              & $\rightarrow$ & C$_{5}$N + H$_2$   &  8$\times10^{-8}$      &  -0.7  &  0.0  &   10-300  &  4\%  & 8$\times10^{-8}$      &  -0.7  & 0.0  & 8$\times10^{-8}$      &  -0.7  & 0.0 \\
8  & & H$_3$C$_{5}$N${^+}$ + $e^-$  & $\rightarrow$ & HC$_{5}$N + H$_2$  &  1$\times10^{-6}$      &  -0.3  &  0.0  &   10-300  & 50\%  & 1$\times10^{-6}$      & -0.3   & 0.0  & 1$\times10^{-6}$      & -0.3   & 0.0  \\
   & &                              & $\rightarrow$ & C$_{5}$N + H$_2$+H &  1$\times10^{-6}$      &  -0.3  &  0.0  &   10-300  & 50\%  & 1$\times10^{-6}$      &  -0.3  & 0.0  & 1$\times10^{-6}$      & -0.3   & 0.0 \\
    \hline
	\end{tabular}
	\end{center}
\end{table}
\end{landscape}                                                             


Note that the experimental determination of $k(T)$ has been done for temperatures as low as 24~K. Therefore, the values at lower temperatures rely on the fitting procedure. The fit, however, can be significantly different from the real values (the rate coefficient extrapolation at temperatures as low as 10~K is not a straightforward procedure. 
Given the differences between our calculated rate coefficients and the fit of the experimental points at 10~K, we ran the model using two networks: a first network in which all the studied reactions are considered with the theoretical values (which corresponds to the network of Tab.~\ref{tab:revised-network}), and a second network in which for the reactions [1], [2] and [3] we adopted the fitted experimental values. In both cases, the reactions [4a], [4b] and [5] are assumed with the theoretical values derived in this work. These results are also compared to the predictions obtained using the original network. 

We modeled two cases with different conditions: a cold (10 K) prestellar core, described in Sec.~\ref{subsec:astro-model-cloud}, and a warm (90 K) protostellar molecular shock, described in Sec.~\ref{subsec:astro-model-shock}. 
The model is briefly described here, while a detailed description can be found in \cite{giani2023revised}. 

The results are described in Secs.~\ref{subsubsec:modelresults-mc} and \ref{subsubsec:astro-shock-results}, and the comparison between the predicted abundances and those observed, described in Sec.~\ref{sec:observations}, is discussed in Sec.~\ref{subsec:astro-compa}.

\subsection{Cold molecular cloud} \label{subsec:astro-model-cloud}

\subsubsection{Modeling and adopted parameters}

In order to simulate a cold molecular cloud, we assumed a gas temperature $T=10$ K and an H-nuclei density $N_H=2\times 10^4$~cm$^{-3}$. Dust grains, which are taken into account to allow the formation of H$_2$, are considered to have radius $a_d$=0.1 $\mu$m and grain density 3~g~cm$^{-3}$, with a gas-to-dust ratio in mass equal to 0.01. The initial elemental abundances are reported in Tab.~\ref{tab:astro-cold-initial-abd}.
We assumed a cloud with A$_v$=100 mag and with a cosmic-ray (CR) ionization rate $\zeta_{\rm CR}=1\times 10^{-17}$ s$^{-1}$. 
We also ran a small grid of models with visual extinction A$_v$ between 100 and 2 mag, cosmic-ray ionisation rate $\zeta_{\rm CR}$ between $1\times 10^{-17}$ and $1\times 10^{-16}$~s$^{-1}$ and C/O ratio between 0.77 and 1.1, in order to check the effects of these parameters on the predicted abundance of HC$_5$N. 

\begin{figure}
\includegraphics[width=8.5cm]{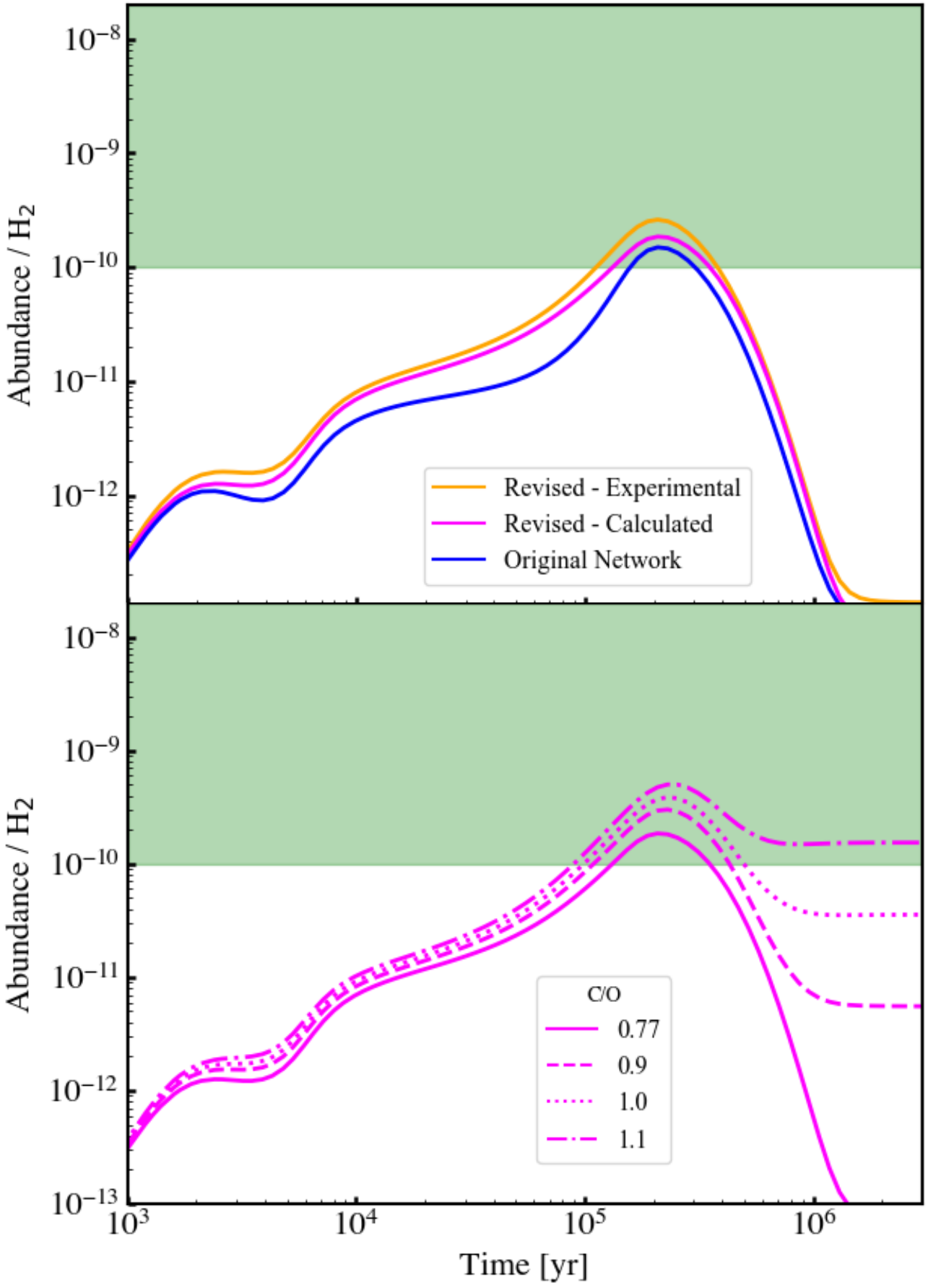}
    \caption{Top panel: HC$_5$N abundances with respect to H$_2$ obtained with the network proposed in this work, using the theoretical results (magenta) ad the experimental results (orange), and the original network (blue). The green band shows the observed abundances of HC$_5$N in L1544 \citep{bianchi2023cyanopolyyne}, obtained assuming an H$_2$ column density of 10$^{22}$ cm$^{-2}$ \citep{Vastel2019}. Bottom panel: HC$_5$N abundances with respect to H$_2$ obtained for initial C/O ratio equal to 0.77 (solid), 0.9 (dashed), 1.0 (dotted) and 1.1 (dotted-dashed). }
    \label{fig:model-MC}
\end{figure}


\subsubsection{Results of the modeling} \label{subsubsec:modelresults-mc}

Figure \ref{fig:model-MC} shows the HC$_5$N abundance as a function of time, obtained using the revised (theoretical and experimental) and the original reaction networks. The HC$_5$N abundance shows a peak at around a few 10$^5$ years, which is related to the availability of carbon atoms before being locked into CO. The HC$_5$N abundance, on the other hand, is found to vary by a factor of 1.4 between the three networks, with the revised experimental network giving the larger HC$_5$N abundance, followed by the revised theoretical network and then by the original one. 

We found that the formation of HC$_5$N is not dominated by a single reaction.
Reaction [5] is one of the dominant reactions in all three model predictions. 
When using the revised networks, the second most important reaction, with a contribution comparable to reaction [5], is reaction [2] (when using the set of experimental values) and reaction [1] (when using the set of theoretical values). In both cases, reaction [4a] is also contributing by 1/4 to the formation of HC$_5$N. 
When using the original network, the reactions [1], [3], and [5] contribute equally to the formation of HC$_5$N.

Figure \ref{fig:model-MC} also shows the effect of different C/O ratios set at the beginning of the simulation. 
The abundance of carbon-chain molecules (cyanopolyynes belonging to this category) is favored by a larger availability of carbon. Thus, as expected, a larger C/O ratio causes an increase of the abundance of HC$_5$N by a factor 5 when going from C/O=0.77 to C/O=1.1.  
We also verified the effect of cosmic rays and visual extinction on the predicted HC$_5$N abundance. A larger value of $\zeta_{\rm CR}$ causes a shift of the abundance at lower times but does not alter the height of the peak. Lower values of visual extinction (which correspond to a larger penetration of the UV) cause a drop in the abundance of HC$_5$N, which is more efficiently destroyed.

\subsection{Warm shocked regions} \label{subsec:astro-model-shock}



\subsubsection{Modeling and adopted parameters}\label{subsubsec:astro-shock}

We followed the same two-step procedure adopted in previous work of our group \citep{podio2014molecular,podio2017silicon,codella2017seeds,codella2020seeds,tinacci2023-gretobape, giani2023revised}. 
Briefly, in the first step, we compute the steady-state abundances in the cloud using $T=10$ K, $N_H=2\times 10^4$~cm$^{-3}$, A$_v$=100~mag and $\zeta_{CR}=1\times 10^{-17}$~s$^{-1}$. 
In the second step, we simulate the passage of the shock by increasing the temperature and density of the gas to $T=90$~K and $N_H=8\times 10^5$~cm$^{-3}$, respectively, and injecting into the gas phase the major components of the grain mantles (see Tab.~\ref{tab:model-parameters-shock}).
We then follow the gas evolution for 10$^4$ years, the time after the shock passage.

\subsubsection{Results of the modeling} \label{subsubsec:astro-shock-results}

\begin{figure}
\includegraphics[width=8.5cm]{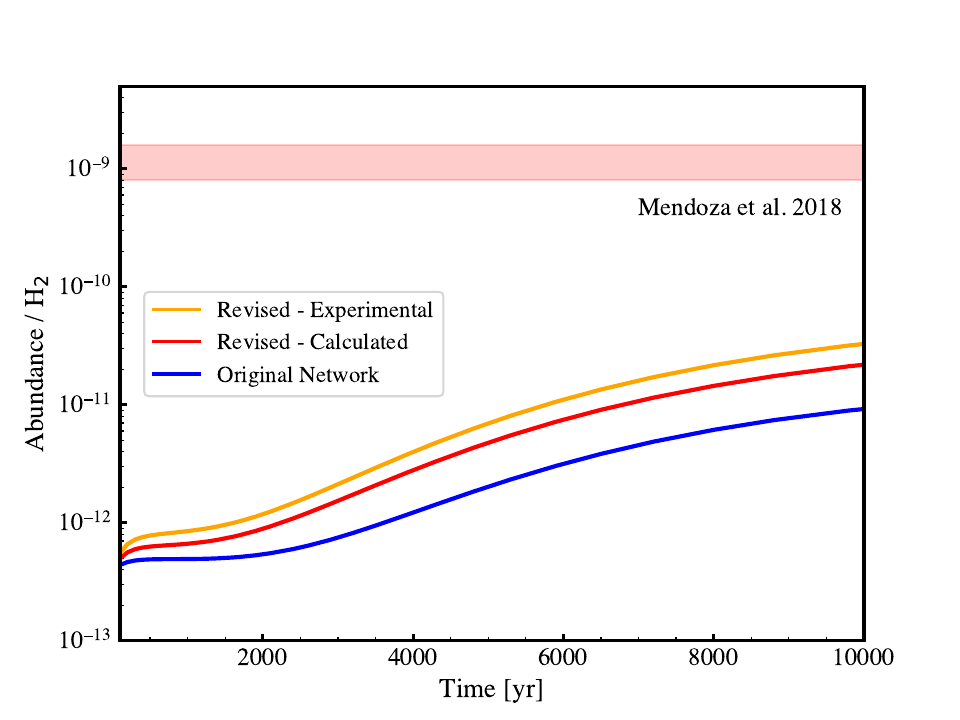}
    \caption{HC$_5$N abundances with respect to H$_2$ obtained with the network proposed in this work, using the theoretical results (red) and the experimental results (orange), and the original network (blue). The red band shows the observed abundances of HC$_5$N in L1157-B1 \citep{2018Mendoza} (see Sec.~\ref{sec:observations} for the details). }
    \label{fig:model-shock}
\end{figure}

Figure \ref{fig:model-shock} shows the post-shock evolution of the gas-phase abundance of HC$_5$N as a function of time, obtained using the revised network, theoretical and experimental, and the original network.

The HC$_5$N abundance does not change significantly during the first 2000 years of the simulation, with an HC$_5$N abundance that remains below 10$^{-12}$ with all the networks tested. After 2000 years, the HC$_5$N abundance increases until at 10$^4$ years, it reaches a few 10$^{-11}$ when using the revised network. The increase is mainly due to reaction [1], which becomes the most important reaction for the formation of HC$_5$N in all three networks. Reaction [1] is favoured by the higher temperatures (rate coefficient larger at 90~K with respect to 10~K) and larger abundance of C$_3$N.

\section{Discussion} \label{sec:discussion}

\subsection{Model predictions versus observations}\label{subsec:astro-compa}

\subsubsection{Cold environment: L1544}\label{subsubsec:tmc-1}

The observations of HC$_5$N in the L1544 prestellar core are described in Sec.~\ref{subsec:obs-results}.
Figure \ref{fig:model-MC} shows the predicted HC$_5$N abundance as a function of time compared to that observed towards L1544. 
The predictions obtained with the three networks agree with the lowest estimate of the measured HC$_5$N abundance. 
Note that the observed HC$_5$N abundance has a large uncertainty due to the uncertainty on the H$_2$ column density. 
Thus, even if our model prediction is not in contradiction with the observations, we cannot assess if it is in agreement or not.
An estimate of the H$_2$ column density in the region of emission of cyanopolyynes is needed to better constrain the HC$_5$N abundance in L1544. 

HC$_5$N is also detected in the TMC-1 molecular cloud \citep{cernicharo2020discovery,loomis2021investigation,cabezas2022discovery} and other cold cores (see Table 3 in \citealt{bianchi2023cyanopolyyne}), with column densities varying between $2.9\times10^{12}$ and $2\times10^{14}$ cm$^{-2}$. However, also in these cases the H$_2$ column density is not well constrained leading to large uncertainties on the HC$_5$N observed abundances.

\subsubsection{Warm environment: L1157-B1}\label{subsubsec:l1157b1}

The observations of HC$_5$N in the L1157-B1 shocked regions are described in Sec.~\ref{subsec:obs-results}.
Figure \ref{fig:model-shock} shows the HC$_5$N abundance predicted by the model using the revised and original networks (see Sec.~\ref{subsubsec:modelresults-mc}), against the observations towards L1157-B1. The model predictions underestimate the observed abundance by almost two orders of magnitude with all three networks. 
This suggests that, while in cold objects neutral-neutral reactions are sufficient to reproduce the HC$_5$N abundance, in warm objects ion-neutral reactions are also needed to explain its formation. This is consistent with the fact that in shocked regions, the lower density and the larger $\zeta_{CR}$ cause a larger presence of ions which might explain the efficient formation of cyanopolyynes.
Moreover, the fact that in the hot corino IRAS 16293–2422 HC$_5$N is observed with a very low abundance in the inner region where ions are less abundant \citep{aljaber2017history}, suggests that different chemistry might be at play in different regions, with ion-neutral or neutral-neutral reactions being dominant depending on the physical conditions (density and $\zeta_{CR}$).



\subsection{Theoretical and experimental rate coefficients: differences on the predicted HC$_5$N abundance}\label{subsubsec:comp-exp-theo}

Figures \ref{fig:model-MC} and \ref{fig:model-shock} show that there are no significant differences in the model prediction of cold and warm regions obtained using the theoretical and experimental rates for reactions [1], [2] and [3]. 
The similarity between the results is due to the compensation of the different rate coefficients used. At 10 K, the fit of the experimental rate coefficient of reaction [1] is one order of magnitude lower than the rate derived for the same reaction from the QM calculations. At the same time, the fit of the experimental rate coefficient of reaction [2] is one order of magnitude larger than the rate derived from QM calculations. The compensation between these two reactions, together with reaction [5] being one of the most important, causes the model prediction obtained with the two networks to be almost the same. 

The comparison of the model predictions obtained using the revised networks with the original one shows again very small differences. This is due to the fact that in the original network, two of the most important reactions for the formation of HC$_5$N, the reactions [1] and [2], have rate coefficients at 10~K quite similar to the theoretical ones. Moreover, the absence of reactions [4a] and [4b], together with a lower value of the rate of reaction [5], cause the model prediction obtained with the original network to be slightly lower than the other two predictions. 

Unfortunately, because of the small differences in the model predictions obtained with the theoretical and experimental results, it is not possible to use the astrochemical model as a tool to discriminate between the different rates, as done in \cite{balucani2024can}.

\subsection{Possible constraints from the isotopic $^{12}$C/$^{13}$C}\label{subsubsec:Cisotopologurs}

\begin{figure}
	\includegraphics[width=8.3cm]{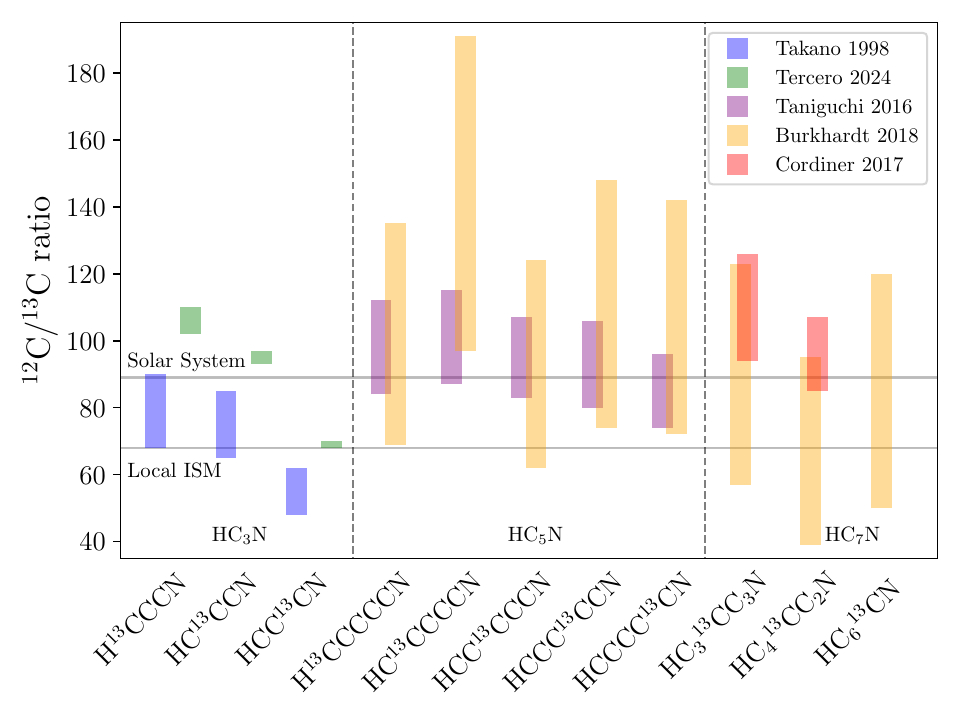}
 	\includegraphics[width=8.7cm]{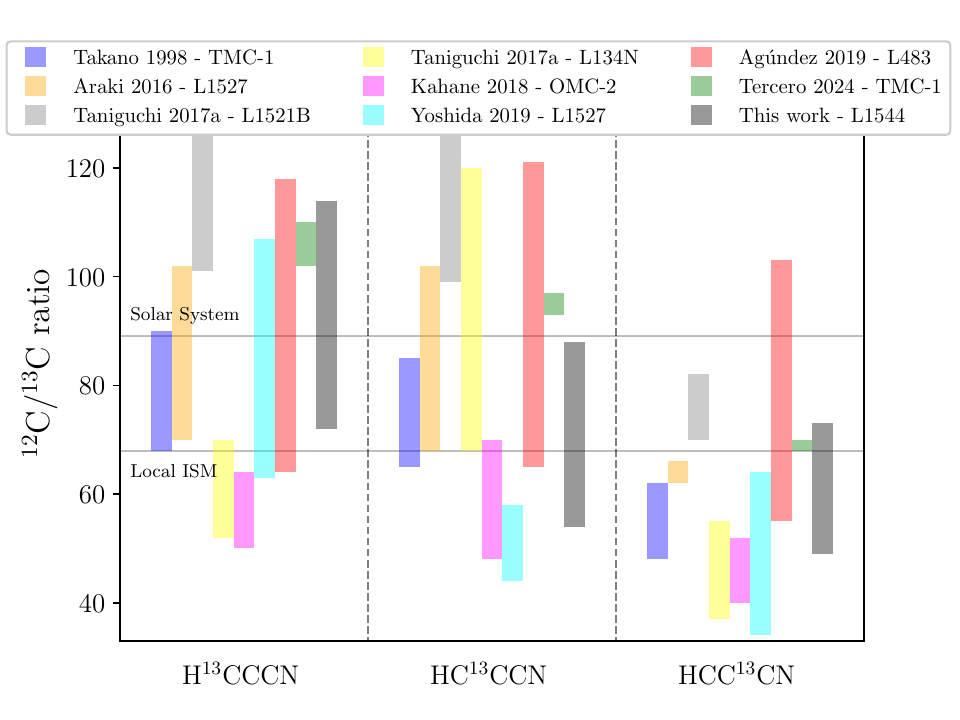}
    \caption{Top panel: Measured $^{12}$C/$^{13}$C ratios of HC$_3$N, HC$_5$N and HC$_7$N in the TMC-1 (CP) molecular cloud \citep{1998Takano,tercero2024doubly,taniguchi2016implication,burkhardt2018detection,cordiner2017deep}. Bottom panel: Measured $^{12}$C/$^{13}$C ratios of HC$_3$N in the TMC-1, L1527, L1521B, L134N, OMC-2 and L483 molecular clouds \citep{1998Takano,araki2016precise,taniguchi201713c,kahane2018first,yoshida2019unbiased,tercero2024doubly}. The grey lines correspond to the $^{12}$C/$^{13}$C ratio measured in the Solar System ($\sim$89) and in the local ISM ($\sim$68) \citep{wilson1994abundances,ritchey2011interstellar}.}
    \label{fig:12C-13C-obs}
\end{figure}                                                                                            

Isotopic fractionation has been used to try to discriminate what are the most important reactions of the formation of cyanopolyynes. The various $^{13}$C isotopologues of HC$_3$N have been observed in the cyanopolyynes peak (CP) region of TMC-1 by \cite{1998Takano} and \cite{tercero2024doubly} and only HCC$^{13}$CN shows a lower $^{12}$C/$^{13}$C ratio (see Fig.~\ref{fig:12C-13C-obs}). \cite{1998Takano} discussed the possible sources of this fractionation considering the formation routes of HC$_3$N. They first concluded that the fractionation occurs during the formation of HC$_3$N and not later through isotope exchange reaction, since the small differences between the ZPE of the three isotopologues (H$^{13}$CCCN, HC$^{13}$CCN and HCC$^{13}$CN) would lead to the same fractionation. 
Ionic pathways involving the HC$_3$NH$^+$ ion are also excluded since its various formation pathways would not lead to selective fractionation. 
Fractionation is instead consistent if HC$_3$N is formed by reactions (i) or (vi) where CN and HNC are $^{13}$C enriched and the $^{13}$C-N bonds are preserved in HC$_3$N. 
Since the relative contribution of reaction (i) or (vi) depends on the HNC/CN abundance ratio, which in turn depends on the age of the cloud, it is possible that cyanopolyynes are formed differently during the evolution of the cloud \citep{taniguchi201713c}. A different formation mechanism of cyanopolyynes depending on the chemical evolution of the cloud might also explain the differences on the $^{12}$C/$^{13}$C ratios observed in more or less evolved clouds. 
So far, TMC-1 and L1521B seem to be the only sources with clear evidence of a larger fractionation of HCC$^{13}$CN with respect to the other $^{13}$C isotopologues of HC$_3$N, while for other cold sources, such as L1527, OMC-2, L134N and L483 the trend seems to be less evident and observations of the same source lead to inconsistent results \citep{araki2016precise,kahane2018first,yoshida2019unbiased,agundez2019sensitive}. Moreover, we found that also in L1544 there is not a clear evidence of a larger fractionation of HCC$^{13}$CN with respect to the other isotopologues (see Sec.~\ref{subsec:obs-results}). 


For HC$_5$N and HC$_7$N, on the other hand, all the various $^{13}$C isotopologues seem to show the same fractionation \citep{burkhardt2018detection,taniguchi2016implication,gratier2016new}.
This suggests that CN is not a major precursor for the formation of cyanopolyynes larger than HC$_3$N and that the reaction of N with hydrocarbon ions, such as C$_5$H$_3^+$ (or C$_7$H$_3^+$) + N, might be the main formation routes of HC$_5$N (or HC$_7$N). 

Considering the large uncertainties on the measured abundances of the HC$_5$N isotopologues and the limited number of observations, we concluded that the $^{12}$C/$^{13}$C ratio does not help to constrain the chemistry behind its formation.
Nevertheless, the $^{12}$C/$^{13}$C observations are consistent with our model predictions, which showed that neither the reaction with CN (reaction [3]), nor the reactions with HC$_3$N are dominant for the formation of HC$_5$N (see Sec.~\ref{subsec:astro-model-cloud}). Indeed, HC$_5$N is not formed by the smaller cyanopolyynes member HC$_3$N, but its formation is instead the result of several reactions with similar contributions, with the reaction of the polyynes member C$_6$H with nitrogen being one of the most important. The fact that cyanopolyynes larger than HC$_3$N appear to derive from the polyynes and not from the smaller cyanopolyynes might explain why the fractionation of the carbon atom of the cyano group is not inherited from HC$_3$N to the larger cyanopolyynes.

\section{Conclusions} \label{sec:conclusions}

In this work, we revised the gas-phase formation of HC$_5$N. Nine gas-phase reactions have been considered as major routes and are in one or both major astrochemical databases, namely KIDA and UDfA (see Sec. 
\ref{sec:overview-gas-routes} and  in Fig. \ref{fig:reac-scheme}). 
We calculated the potential energy surfaces using a combination of DFT and CCSD(T) methods while, for the kinetic, we used capture and RRKM theories for six of them. 
Moreover, we report the experimental results obtained with the CRESU technique for three of the six revised reactions. More specifically:
\begin{itemize}
    \vspace{-7pt}
    \item[(i)] For three neutral-neutral reactions (C$_3$N + C$_2$H$_2$, C$_2$H + HC$_3$N, and CN + C$_2$H$_4$), we measured the global rate coefficients as a function of the temperature (the lowest T investigated is ca. 24 K), performed electronic structure calculations of the underlying potential energy surfaces and determined the rate coefficients and product branching fractions. To the best of our knowledge, there are the first measurements at low temperatures and the first PESs ever published (with the exception of the CN + C$_4$H$_2$ reaction for which a previous PES has been determined).  
    Experimental and theoretical results are in general in good agreement, with the largest difference being a factor 6. Since the CRESU technique in the PLP arrangement cannot provide the product branching fractions, the experimental and theoretical results nicely complement each other. We recall that in astrochemical models, the nature of the products and their BF are important parameters since the products of one reaction will be the reactants of a subsequent one in the intricate network of reactions that are necessary to account for the global chemistry;
    \item[(ii)] 
    For three other reactions (C$_4$H + HCN, C$_4$H + HNC, and C$_6$H + N), we performed electronic structure calculations of the underlying potential energy surfaces and determined
   the rate coefficients and product branching fractions. To the best of our knowledge, there are the first PESs ever published (in the case of the C$_6$H + N reaction, a partial characterization of the PES was available \citep{loison2014interstellar});  
    \item[(iii)] We did not investigate two dissociative recombination reactions involving the ions HC$_5$NH$^+$ and H$_3$C$_5$N$^+$, as well as various ion-molecule reactions leading to their formation.
\end{itemize}

Our revised reaction network for the formation of HC$_5$N is summarized in Tab.~\ref{tab:revised-network}.
We tested the impact of the revised network on the predicted abundance of HC$_5$N by modeling two representative objects of cold and warm environments: the L1544 prestellar core and the L1157-B1 shocked region, respectively.
We found that the formation of HC$_5$N is the result of the contributions of all the following reactions

\begin{enumerate}
    \item[(1)] C${_3}$N + C$_{2}$H$_{2}$ $\rightarrow$  HC$_{5}$N + H
    \item[(2)] HC${_3}$N + C$_2$H $\rightarrow$  HC$_{5}$N + H
    \item[(4a)] C$_{4}$H + HCN $\rightarrow$  HC$_{5}$N + H
    \item[(5)] C${_6}$H + N $\rightarrow$  HC$_{5}$N + C
\end{enumerate}

In cold regions, the most important contribution is given by reaction [5], followed by reaction [1] and [4a], while, in warm regions, reaction [1] becomes the most important one. 
Please note that reaction [2] is one of the most important in cold regions only if the experimental rate coefficient is adopted.

Reaction [5] was already present in the KIDA database with estimated rate coefficient and BFs, but we updated them following our QM calculations. Our calculated rate coefficient is consistent with that of the database, but the HC$_5$N + C channel is found to be the main one while the channel leading to C$_6$H + N gives a marginal contribution. 
Reactions [1] and [2] were studied both theoretically and experimentally and, depending whether the experimental or theoretical results are adopted, the two results cause one reaction or the other to be important for the formation of HC$_5$N.

For reaction [4a], we adopted our new calculated rate coefficient, which is three orders of magnitude higher than the estimated value reported in the KIDA database.


We verified whether the isotopic $^{12}$C/$^{13}$C ratio could help us to constrain the possible formation routes of HC$_5$N. We searched for the emission of the various isotopologues of HC$_3$N and HC$_5$N in L1544. We detected one transition from each HC$_3$N isotopologue, but not from the HC$_5$N isotopologues. Thus, because of the limited number of observations of the isotopologues of HC$_5$N and the large uncertainties on the existing observations, we concluded that the $^{12}$C/$^{13}$C ratio does not help to constrain the chemistry of HC$_5$N.

Finally, the fact that the model predictions are in agreement with the observed HC$_5$N abundance in L1544 but not in L1157-B1 suggests that ionic chemistry might be particularly relevant in shocked regions for the formation of cyanopolyynes. We plan to revise the ion-neutral reactions in future work.

\section*{Acknowledgements}

This project has received funding within the European Union’s Horizon 2020 research and innovation programme from the European Research Council (ERC) for the project “The Dawn of Organic Chemistry” (DOC), grant agreement No 741002, and from the Marie Sklodowska-Curie for the project ”Astro-Chemical Origins” (ACO), grant agreement No 811312.
N.B., E.B., and M.R. acknowledge support from the Italian Space Agency (Bando ASI Prot. n. DC-DSR-UVS-2022-231, Grant no. 2023-10-U.0 MIGLIORA).
M.R. acknowledges financial support under the National Recovery and Resilience Plan (NRRP), Mission 4, Component 2, Investment 1.1, Call for tender No. 104 published on 2.2.2022 by the Italian Ministry of University and Research (MUR), funded by the European Union – NextGenerationEU– Project Title 2022JC2Y93 ChemicalOrigins: linking the fossil composition of the Solar System with the chemistry of protoplanetary disks – CUP J53D23001600006 - Grant Assignment Decree No. 962 adopted on 30.06.2023 by the Italian Ministry of Ministry of University and Research (MUR). 
Some computations presented in this paper were performed using the GRICAD infrastructure (https://gricad.univ-grenoble-alpes.fr), which is partly supported by the Equip@Meso project (reference ANR-10-EQPX-29-01) of the programme Investissements d'Avenir supervised by the Agence Nationale pour la Recherche. \\
E.B. acknowledges support from the Deutsche Forschungsgemeinschaft (DFG, German Research Foundation) under Germany´s Excellence Strategy – EXC 2094 – 390783311. The Green Bank Observatory is a facility of the National Science Foundation operated under cooperative agreement by Associated Universities, Inc.
E.B. acknowledges contribution of the Next Generation EU funds within the National Recovery and Resilience Plan (PNRR), Mission 4 - Education and Research, Component 2 - From Research to Business (M4C2), Investment Line 3.1 - Strengthening and creation of Research Infrastructures, Project IR0000034 – “STILES - Strengthening the Italian Leadership in ELT and SKA”.\\
MF, SCDE, J-CG and IRS acknowledge funding from the Region of Brittany. This work was supported by the French National Programme “Physique et Chimie du Milieu Interstellaire” (PCMI) of CNRS/INSU with INC/INP co-funded by CEA and CNES.

\section*{Data availability}
The data underlying this article will be shared on reasonable request to the corresponding author.


\bibliographystyle{mnras}
\bibliography{LisaGiani}


\appendix

\section{Detailed Experimental Results} \label{sec:appendix-results}


\subsection{Specific methods and results for reaction [1]: \texorpdfstring{C${_3}$N + C$_{2}$H$_{2}$  $\rightarrow$  HC$_{5}$N + H}{reaction1} \label{subsec:append-exp-R1}}

The kinetics of reactions involving the C$_3$N radical have not been studied previously, to the best of our knowledge. We therefore give here a comprehensive description of the production and detecton schemes. BrC$_3$N was chosen as a photolytic precursor by analogy with CN radicals which can be produced by laser photolysis of BrCN as in \cite{RN1222}.
The synthesis of the precursor was adapted from the description available in the literature. Under the protocol of Moureu and Bongrand \citep{RN1217} propiolamide is first synthesised by adding methyl propiolate to liquid ammonia under a neutral atmosphere. 82 g of solid ammonia had previously been obtained by flowing gaseous ammonia into a liquid nitrogen-cooled flask. The flask is then heated up to the point where ammonia melts (by immersion in a -60 $^\circ$C methanol bath). 50 g of methyl propiolate is then added dropwise to the solution and the reactive medium is agitated under nitrogen flow for 30 minutes. Low boiling compounds are removed at room temperature under vacuum to give pure propiolamide at 90\% yield.
The next step is the synthesis of cyanoacetylene. 20 g of crushed propiolamide is mixed with 60 g of sand and 60 g of P$_4$O$_{10}$. The mixture is heated up to 200 $^\circ$C over the course of three hours and the product is captured in a cold finger (liquid N$_2$) after flowing through a trap at $-$25 $^\circ$C. The product is obtained as the form of white crystals with 63\% yield.

The last step is the synthesis of bromocyanoacetylene itself from the description of Kloster-Jensen \citep{RN1102}. A solution of KBr/Br$_2$ is prepared by adding 3 mL of Br$_2$ to a solution of KBr (prepared by dissolving 7.5 g of KBr in 50 mL of cold water). To this solution is added a solution of cyanoacetylene (1.2 g in 45 mL of water). After mixing the two solutions, a half normal KOH solution is added slowly and once added, the mixture is stirred until the precipitation of BrC$_3$N appears. The mixture is then filtered and dried by pumping through a P$_4$O$_{10}$ solid half filed tube and obtained at low temperature in a cold finger (liquid N$_2$).
Properties measured for the product include a tendency to decompose when exposed to light on a long-term basis (over a week), a highly hygroscopic, a strong lachrymator effect and a tendency to adsorb on metallic surfaces. The experimental vapour pressure is 26.0 mbar at 19 $^\circ$C, measured directly after separation via a calibrated absolute pressure gauge. The synthesis itself presents no particular safety challenge (except the use of a bromine solution).
The BrC$_3$N can be stored under vacuum in a glass vessel on a relatively long-term basis without any problem, as long as it is kept at low temperature (4$^\circ$C) and not exposed directly to light. Depending on the quality of the trap keeping it, it may be necessary to dry it again (by flowing it through solid phosphorous pentoxide) before use.

\begin{figure*}
 	\includegraphics[width=13cm]{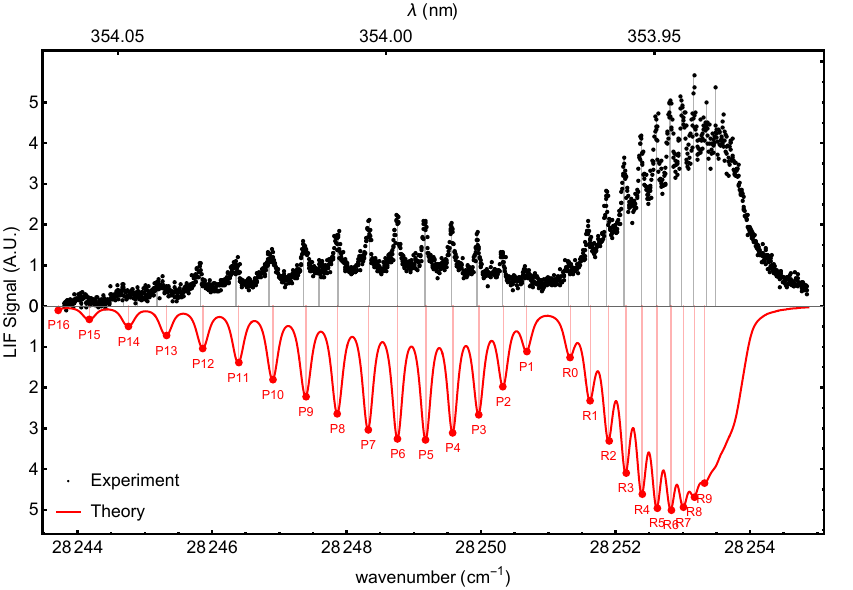}
    \caption{LIF excitation spectrum for the $^2\Sigma^+$--$^2\Sigma^+$ band of C$_3$N within the $\mathrm{\tilde{B}}^2\Pi_i$--$\mathrm{\tilde{X}}^2\Sigma^+$ system, obtained at a pump-probe delay time of 30 $\mu$s at 15.5 K, compared to a PGOPHER simulation \protect\cite{RN1221} using spectroscopic parameters from \protect\cite{RN1220}, with Lorentzian peaks of FWHM 0.22 cm$^{-1}$.}
    \label{fig:lif}
\end{figure*}

For further use in the experiment, small amounts of C$_3$N were sublimated into a $\sim$15 L tank up to a few mbar of total pressure. The tank was then filled with pure helium up to $\sim$2 bar. A large number of these mixtures were realised during these experiments and no degradation was detected with the BrC$_3$N in gas phase. This gas mixture was injected in the system along with the carrier gas, via its own flow controller (MKS), at a flow rate of $\sim$30 standard cm$^3$ min$^{-1}$.
The C$_3$N is obtained by photolysis at 248 nm initiated by a KrF excimer laser (LPX 210i, Lambda Physik, now Coherent Inc.). Typical laser fluences were 50 mJ cm$^{-2}$, and no difference in the kinetics results was detected when doubling or halving this value. The photolysis laser beam was directed counter flow, entering through a Brewster angle quartz window and traversing the whole chamber up to the other side of the CRESU setup.
C$_3$N was detected by the laser-induced fluorescence (LIF) method. An extensive study by Hoshina and Endo \citep{RN1220} measured vibronic bands between 27000 cm$^{-1}$ (370 nm) and 29400 cm$^{-1}$ (340 nm). We used a pulsed dye laser system to reach wavelengths between 350-370 nm. This system is composed of a Nd:YAG laser (Continuum, Powerlite Precision II 9000) pumping a Sirah Cobra Stretch pulsed dye laser (double 1800 lines/mm gratings), operating with Pyridine-2 dye (Exciton, Inc) and further doubled with a KDP nonlinear crystal, for UV generation. The system was calibrated with the help of a wavelength meter (High Finesse WS/7L), and a linewidth of 3 pm (in the UV) was measured. The laser was propagated along the flow, passing through two Brewster angle quartz windows (respectively interfaces between exterior/chamber, and chamber/reservoir). Laser fluences as high as 20 mJ cm$^{-2}$ could be reached, but a better signal/noise was achieved at lower energies, $\sim$2 mJ cm$^{-2}$, since at high fluence, some noise was originating from diffuse light inside the system. The detection of signal took place via a photomultiplier tube (PMT) (Electron Tubes model 9125QB), positioned on the topside of the apparatus. An interference filter positioned before the PMT (Newport 10BPF10-420) allowed the collection of fluorescence light out of resonance, achieving a better signal/noise ratio. A typical fluorescence lifetime of (110 $\pm$ 20) ns was observed, similar to Hoshina and Endo’s reported value of (110 $\pm$ 10) ns. A blue-tinted glass window was also used as a low-pass filter to remove some additional light from both laser pulses and to act as a window against the pressure between the PMT (operating in air) and the chamber.
The system was synchronized by means of a delay generator (Stanford Research Systems DG535), which sent trigger pulses both to the excimer laser and the dye laser’s pump laser. The signal measured by the PMT is passed to a gated integrator (Stanford Research Systems SR250), which receives a synchronisation signal from the DG535. The integrated value of the fluorescence signal is then passed to a Labview program via an SR 245 (Stanford Research Systems) interface. The purpose of this program is to manage all instrument signals and process the data. In particular it controls the gas flows, crucial for the rate coefficient measurement, and changes the delays between lasers or their wavelength.
To ensure the proper detection of C$_3$N, vibronic bands were observed at different pump-probe time delays and assigned to the system $\mathrm{\tilde{B}}^2\Pi_i$--$\mathrm{\tilde{X}}^2\Sigma^+$.
The band centred at 353.97 nm (28251 cm$^{-1}$) was identified as the $^2\Sigma^+$--$^2\Sigma^+$ band. 
Figure \ref{fig:lif} shows a fluorescence excitation spectrum obtained at 15.5 K compared to a PGOPHER simulation \citep{RN1221} using spectroscopic parameters from \cite{RN1220}, with Lorentzian peaks of FWHM 0.22 cm$^{-1}$, chosen to fit best the experimental spectrum. Subsequent measurements took place at each temperature used in the rate coefficient measurements, and the wavelength used was where the intensity would be the highest at that temperature.

The co-reagent, acetylene (C$_2$H$_2$, Air Liquide 99.7\%) was purified with an active charcoal system (Matheson Gas purifier model 450B and cartridge 454) to remove traces of acetone in which it is stabilised for storage. The acetylene and diluted BrC$_3$N gases were flowed into the main carrier gas flow (He  99.995\%, N$_2$ 99.995\%, Ar 99.998\%, Air Liquide) via calibrated mass flow controllers (MKS Instruments).

Table \ref{tab:k-reac-1} presents the data, with the temperature of the measurements, the carrier gas (argon, helium or nitrogen), the flow density and range of reagent densities, how many experimental points were measured for each second order plot, and rate coefficients with their associated uncertainties. The latter were determined statistically at the 95\% confidence level and combined in quadrature with a likely 10\% systematic error to yield the quoted values.

\begin{table}
	\begin{center}
 \renewcommand{\arraystretch}{1.4}
  \setlength{\tabcolsep}{3.5pt}
 \caption{rate coefficients $k_{2nd}$ measured for reaction [1] at different temperatures in different carrier gases M. Errors are calculated as 95\% statistical uncertainty combined in quadrature with an estimated 10\% likely systematic error.  Where more than one measurement was made at a single temperature, these values are combined as a weighted average (with weighting undertaken before addition of the estimated likely systematic error).}
\label{tab:k-reac-1}
\begin{tabular}{cccccc}
\hline
 T  &  M   &  [M]   &  [C$_4$H$_2$]   &  Number   &   $k_{2nd}$ \\
 (K) &   & (10$^{16}$ cm$^{-3}$)  &  (10$^{13}$ cm$^{-3}$)  & of points  &   (10$^{-10}$ cm$^{-3}$ s$^{-1}$) \\
\hline
\hline
  24.1  &  He   &  18.3  &  0-3.187  &  10  &  3.72 $\pm$ 0.39  \\
  \hline
 39     & N$_2$ &  3.21  &  0-4.179  &  10  &  3.52 $\pm$ 0.36  \\
 \hline
  52.3  &  Ar   &  10.3  &  0-1.590  &  10  &  4.39 $\pm$ 0.46  \\
  \hline
  82.6  & N$_2$ &  4.88  &  0-1.107  &  11  &  4.67 $\pm$ 0.47  \\
  \hline
  112   &  Ar   &  2.78  &  0-4.365  &  9   &  4.56 $\pm$ 0.47  \\
  \hline
  200   & N$_2$ &   5.57 &  0-6.750  &  10  &  3.38 $\pm$ 0.35  \\
  \hline
  294   & N$_2$ &  29.08 &  0-25.36  &  8   &  2.51 $\pm$ 0.27  \\
  294   & N$_2$ &  7.520 &  0-15.43  &  12  &  2.47 $\pm$ 0.26  \\
        &       &        &           &      &  2.48 $\pm$ 0.28  \\
\hline
\end{tabular}\\
	\end{center}  
\end{table}

The experimentally determined rate coefficients were fitted to a modified Arrhenius or Kooij expression yielding the result:

\begin{equation}
    k_{\text{C${_3}$N + C$_{2}$H$_{2}$}}(T)=3.091\times10^{-10}{\left(\frac{T}{300}\right)}^{-0.577}\exp{\left(\frac{-33.64}{T}\right)}~~~ \text{cm}^3 \text{s}^{-1}
\end{equation}

An average (rms) deviation of this expression from the experimental points of $4.13\times10^{-11}$ cm$^3$ s$^{-1}$ can be calculated, to which should be added in quadrature an estimated likely systematic error of 10\%.
The rapidity of this reaction between 24 and 294 K and the negative dependence of its rate coefficient on temperature imply that the reaction is exothermic and barrierless. To our knowledge, no previous measurements of the rate coefficient exist at any temperature.

\subsection{Specific methods and results for reaction [2]: \texorpdfstring{C$_2$H + HC$_{3}$N $\rightarrow$ HC$_{5}$N + H}{reaction3} \label{subsec:append-exp-R2}}

In contrast to the C$_3$N studies, the kinetics of C$_2$H radical reactions have been previously studied by the CRESU technique \citep{RN638,RN648}, and so only a very brief description will be given here. C$_2$H radicals were generated by pulsed laser photolysis of acetylene, C$_2$H$_2$ (source and purity as noted in the previous section) or trifluoropropyne, CF$_3$C$_2$H (ABCR 98\%, used diluted at 10\% in He). The photolysis was achieved at 193 nm using an ArF excimer laser (Tui Laser, ExciStar 200-XS-T), operating at a repetition rate of 10 Hz and a pulse duration of $\sim$10 ns. The laser fluence used was 15 mJ cm$^{-2}$. The progress of C$_2$H reactions was monitored by observing chemiluminescence from CH ($\mathrm{A}^2\Delta^+$), which is formed as a minor product of the reaction between C$_2$H and O$_2$, which is flowed in along with the other gases. Details of this method and possible complications are discussed at length by Chastaing et al. \citep{RN648}. One of its disadvantages is that, to study any reaction of C$_2$H, it is necessary to have both C$_2$H$_2$ and O$_2$ present, one as the photochemical precursor of the radicals and the other to generate the chemiluminescent marker, CH ($\mathrm{A}^2\Delta^+$), as well as any other reagent such as HC$_3$N. The ‘background rate’, i.e. that with no HC$_3$N, is substantial and means that relatively large concentrations of any third reagent must be added in order to change the first-order decay rates appreciably. This ‘background rate’ was found to be somewhat smaller and chemiluminescence signals somewhat larger using the (more expensive) trifluoropropyne precursor.
The HC$_3$N co-reagent was synthesised in molar quantities as needed according to the same procedure as detailed above for the initial steps of the synthesis of BrC$_3$N. The procedures for its storage, characterisation and use were identical to those described in our previous study of the reaction of CN radicals with HC$_3$N  \citep{RN600}. The purity was measured to be 99\%.  
Chemiluminescence signals were detected by a photomultiplier (Thorn EMI, 6723 QL30F) after passing through a 10 nm FWHM interference filter centred at 428 nm and then digitised by an acquisition card (National Instruments PCI 6229), before being recorded on a computer for averaging and subsequent analysis. Each decay was typically the result of 500 to 1000 averages. These traces were fitted to single exponential functions, yielding pseudo-first-order rate coefficients related to the rate of loss of C$_2$H radicals. These were recorded for a range of different HC$_3$N concentrations, specified in Table \ref{tab:k-reac-2}, enabling the construction of second-order plots and the determination of the bimolecular rate coefficient at each temperature. At most temperatures, more than one determination was undertaken, and the results are combined in the table as a weighted average (using the statistical uncertainties only for the weighting).

\begin{table*}
	\begin{center}
 \renewcommand{\arraystretch}{1.4}
 \caption{rate coefficients $k_{2nd}$ measured for reaction [2] at different temperatures in different carrier gases M. Errors are calculated as 95\% statistical uncertainty combined in quadrature with an estimated 10\% likely systematic error.  Where more than one measurement was made at a single temperature, these values are combined as a weighted average (with weighting undertaken before addition of the estimated likely systematic error).}
\label{tab:k-reac-2}
\begin{tabular}{ccccccccc}
\hline
 T (K)  &  M   &  [M]  &  C$_2$H   &  [Precursor]   &  [O$_2$]    &  [HC$_3$N]    &  Number    &   $k_{2nd}$  \\
   &    &  (10$^{16}$ cm$^{-3}$) &   precursor  &   (10$^{13}$ cm$^{-3}$) &   (10$^{14}$ cm$^{-3}$)  &   (10$^{13}$ cm$^{-3}$)  &  of points  &   (10$^{-10}$ cm$^{-3}$ s$^{-1}$) \\
\hline
\hline
  24  &  He   &  18.30 & CF$_3$C$_2$H  & 4.81 & 2.35  &  0.00-4.03  &  12  &  5.61 $\pm$ 0.65  \\
  24  &  He   &  18.30 & CF$_3$C$_2$H  & 4.86 & 3.35  &  0.24-4.63  &  11  &  5.62 $\pm$ 0.75  \\
    &     &   &  &  &   &    &   &  5.61 $\pm$ 0.62  \\
  \hline
 30.7   & Ar &  1.90 & CF$_3$C$_2$H  & 2.61 & 3.51 &  0.00-3.95  &  9  &  3.52 $\pm$ 0.36  \\
 \hline
  52.3  &  Ar   &  10.3  & CF$_3$C$_2$H  & 3.63 & 2.78 &  0.00-4.19  &  11  &  2.71 $\pm$ 0.29  \\
  \hline
  83  & N$_2$ &  4.88  & CF$_3$C$_2$H  & 4.30 & 4.70 &  0.48-4.69  &  7  &  1.68 $\pm$ 0.24  \\
  83  & N$_2$ &  4.88  & CF$_3$C$_2$H  & 4.45 & 1.91 &  0.13-3.42  &  7  &  1.33 $\pm$ 0.20  \\
    &     &   &  &  &   &    &   &  1.48 $\pm$ 0.19  \\
  \hline
  97.4   &  Ar   &  15.45 & CF$_3$C$_2$H  & 3.78 & 1.59   &  0.34-8.85  &  13   &  1.15 $\pm$ 0.13  \\
  \hline
  145.5  & N$_2$ &  9.23  & CF$_3$C$_2$H  & 3.07 & 2.52 &  0.88-25.69  &  11  &  0.85 $\pm$ 0.09  \\
  \hline
  200   & N$_2$ &   5.57 & CF$_3$C$_2$H  & 4.64 & 1.95 &  0.35-7.39  &  7  &  0.54 $\pm$ 0.11  \\
  \hline
  297   & N$_2$ &  45.00 & CF$_3$C$_2$H  & 3.33 & 8.02 &  3.02-63.0  &  9   &  0.47 $\pm$ 0.05  \\
  297   & N$_2$ &  15.40 & C$_2$H$_2$    & 4.93 & 7.58 &  0.05-47.0  &  9   &  0.53 $\pm$ 0.06  \\
  297   & N$_2$ &  15.75 & CF$_3$C$_2$H  & 3.89 & 6.59 &  0.03-41.7  &  9   &  0.49 $\pm$ 0.05  \\
    &     &   &  &  &   &    &   &  0.49 $\pm$ 0.05  \\
\hline
\end{tabular}\\
	\end{center}  
\end{table*}

The experimentally determined rate coefficients showed a marked negative temperature dependence and were fitted to a $T^n$ expression yielding the result:

\begin{equation}
    k_{\text{C$_2$H + HC${_3}$N}}(T)= 3.91\times10^{-11}{\left(\frac{T}{300}\right)}^{-1.04}   ~~~ \text{cm}^3 \text{s}^{-1}
\end{equation}

An average (rms) deviation of this expression from the experimental points of 2.47 $\times$ 10$^{-11}$ cm$^3$ s$^{-1}$ can be calculated, to which should be added in quadrature an estimated likely systematic error of 10\%.
The rapidity of this reaction between 24 and 297 K and the negative dependence of its rate coefficient on temperature implies that the reaction is exothermic and barrierless. To our knowledge, no previous measurements of the rate coefficient exist at any temperature.

\subsection{Specific methods and results for reaction [3]: \texorpdfstring{CN + C$_{4}$H$_{2}$ $\rightarrow$ HC$_{5}$N + H}{reaction3} \label{subsec:append-exp-R3}}

The kinetics of CN radical reactions have been previously studied using the CRESU technique, and the reader is referred to the article by Cheikh Sid Ely et al. for a full description \citep{RN600}. Briefly, CN radicals are produced by the photolysis of cyanogen iodide (ICN; Acros Organics, 98\%) at 266 nm using the fourth harmonic output of a Nd:YAG laser (Spectra Physics GCR 190) at a fluence in the reaction zone of $\sim$35 mJ cm$^{-2}$. They were detected by LIF with excitation in the (0,0) band of the CN ($\mathrm{A}^2\Sigma^+$--$\mathrm{X}^2\Sigma^+$) system at $\sim$388 nm using the output of a tunable dye laser (Continuum ND6000) operating with Exciton Exalite 389 dye, pumped by the tripled output of another Nd:YAG laser (Spectra Physics GCR 230). The photolysis and probe lasers, after combination on a dichroic mirror, entered the CRESU chamber and gas reservoir via two quartz Brewster angle windows and passed through the throat of the Laval nozzle and along the axis of the gas flow. Fluorescence in the (0,1) band at $\sim$420 nm was collected at a certain distance from the Laval nozzle (10-30 cm), depending on the specific Laval nozzle, using a combination of a UV-enhanced mirror and fused silica lenses onto the photocathode of a UV-sensitive photomultiplier tube (Thorn EMI 6723) through a narrow band interference filter centered at 420 nm with a 10 nm FWHM bandwidth (Ealing Optics). 
Diacetylene, C$_4$H$_2$, was synthesised following the procedure of \cite{khlifi1995absolute}. All reagents were obtained from Sigma Aldrich at standard reagent grade purity. In a 250 mL three-necked round bottomed glass vessel, 36 g of potassium hydroxide (KOH) were dissolved in 360 mL of water and 36 mL of tetraethylene glycol dimethyl ether was added; 36.6 mL of 1,4-dichloro-2-butyne (Sigma-Aldrich, 99\% purity) was then added. A continuous flow of nitrogen gas removed all the oxygen from the various parts of the assembly. The mixture was then heated to 100°C using an oil bath and the synthesised diacetylene with traces of water vapour was swept out by the N$_2$ flow and trapped in a liquid nitrogen cooled vessel. Subsequently, the diacetylene was dried by removing the liquid nitrogen and passing the vapour through a P$_4$O$_{10}$ column into a second cold trap, yielding high purity sold diacetylene. The purity of the synthesised C$_4$H$_2$ was determined by infrared spectroscopy to be better than 98\%. As it has a tendency to polymerise at room temperature, it was either kept frozen as a pure solid, or diluted to a few percent in helium in a PTFE coated 13 L pressure vessel, ready for use in the kinetics experiments.
Pseudo-first-order decays were observed by recording the variation in the laser-induced fluorescence (LIF) signal as a function of the time delay between the pulses from the photolysis and probe lasers, for 200 different time delays, averaged typically 10 times. A baseline measurement was also recorded at negative time delays (probe before photolysis pulse) to establish a pre-trigger baseline. These traces of LIF signal versus time delay were fitted to single exponential functions, the fit being started at time delays sufficient to allow for rotational relaxation. This procedure yielded pseudo-first-order rate coefficients related to the rate of loss of CN radicals. These were recorded for a range of different C$_4$H$_2$ concentrations, specified in Table \ref{tab:k-reac-3}, enabling the construction of second-order plots and the determination of the bimolecular rate coefficient at each temperature, as reported in the table.

\begin{table}
	\begin{center}
 \renewcommand{\arraystretch}{1.4}
 \setlength{\tabcolsep}{3.5pt}
 \caption{rate coefficients $k_{2nd}$ measured for reaction [3] at different temperatures in different carrier gases M. Errors are calculated as 95\% statistical uncertainty combined in quadrature with an estimated 10\% likely systematic error.  Where more than one measurement was made at a single temperature, these values are combined as a weighted average (with weighting undertaken before addition of the estimated likely systematic error).}
\label{tab:k-reac-3}
\begin{tabular}{lccccc}
\hline
 T  &  M   &  [M]   &  [C$_4$H$_2$]   &  Number   &   $k_{2nd}$ \\
 (K) &   & (10$^{16}$ cm$^{-3}$)  &  (10$^{13}$ cm$^{-3}$)  & of points  &   (10$^{-10}$ cm$^{-3}$ s$^{-1}$) \\
\hline
\hline
  24.1  &  He   &  18.3  &  0-3.27  &  10  &  4.68 $\pm$ 0.53  \\
  24.1  &  He   &  18.3  &  0-2.97  &  10  &  4.79 $\pm$ 0.52  \\
        &       &        &           &      &  4.75 $\pm$ 0.50  \\    
  \hline
 36     & He &  5.28  &  0-1.17  &  8  &  4.72 $\pm$ 0.56  \\
 \hline
  49.1     & He &  10.40  &  0-3.91  &  9  &  5.07 $\pm$ 0.59  \\
 \hline
  52.3  &  Ar   &  10.3  &  0-2.07  &  7  &  4.53 $\pm$ 0.46  \\
  52.2  &  Ar   &  5.15  &  0-1.70  &  7  &  4.64 $\pm$ 0.47  \\  
        &       &        &           &      &  4.55 $\pm$ 0.46  \\  
  \hline
  71.5  & N$_2$ &  5.79  &  0-4.45  &  9  &  5.21 $\pm$ 0.62  \\
  \hline
  97.4  &  Ar   &  15.45  &  0-2.40  &  8  &  4.72 $\pm$ 0.55  \\
  97.4  &  Ar   &  15.45  &  0-4.22  &  11  &  4.92 $\pm$ 0.50  \\  
        &       &        &           &      &  4.90 $\pm$ 0.50  \\  
  \hline
  145.4   &  N$_2$   &  9.23  &  0-3.33  &  11   &  4.90 $\pm$ 0.50  \\
  \hline
  200   & N$_2$ &   5.57 &  0-3.22  &  10  &  4.32 $\pm$ 0.55  \\
  \hline
  296   & Ar    &  12.00 &  0-8.57  &  8   &  3.78 $\pm$ 0.38  \\
  300   & Ar    &  11.70 &  0-3.61  &  8   &  3.75 $\pm$ 0.41  \\
  293   & N$_2$ &  24.30 &  0-16.2  &  10  &  3.80 $\pm$ 0.39  \\
        &       &        &           &      &  3.79 $\pm$ 0.38  \\
\hline
\end{tabular}\\
	\end{center}  
\end{table}

The experimentally determined rate coefficients were fitted to a modified Arrhenius or Kooij expression yielding the result:

\begin{equation}
    k_{\text{CN + C$_{4}$H$_{2}$}}(T)=  4.06\times10^{-10}{\left(\frac{T}{300~\text{K}}\right)}^{-0.24}\exp{\left(\frac{-11.5}{T}\right)}  ~~~ \text{cm}^3 \text{s}^{-1}
\end{equation}

An average (rms) deviation of this expression from the experimental points of 2.31 $\times$ 10$^{-11}$ cm$^3$ s$^{-1}$ can be calculated, to which should be added in quadrature an estimated likely systematic error of 10\%.
Excellent agreement at room temperature was obtained with the only previously determined value of this rate coefficient (k = (4.20 $\pm$ 0.2) $\times$ 10$^{-10}$ cm$^3$ s$^{-1}$ at 298 K) by \cite{seki1996reaction}. The very high rate coefficients obtained at low temperatures and the slight negative temperature dependence are indicative of a reaction mechanism with no energy barrier.


\begin{figure*}
	\includegraphics[width=10cm]{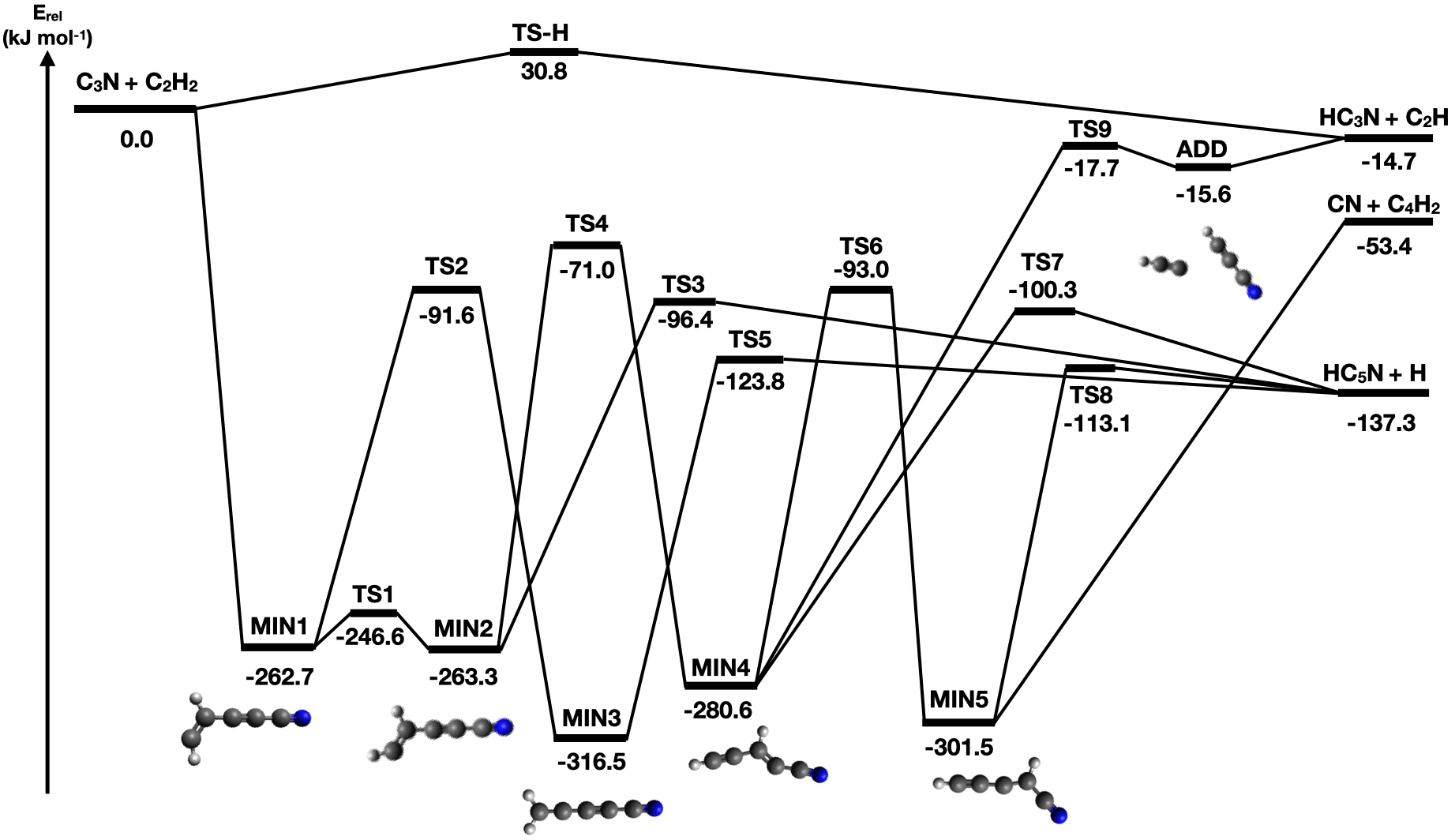}
 	\includegraphics[width=7cm]{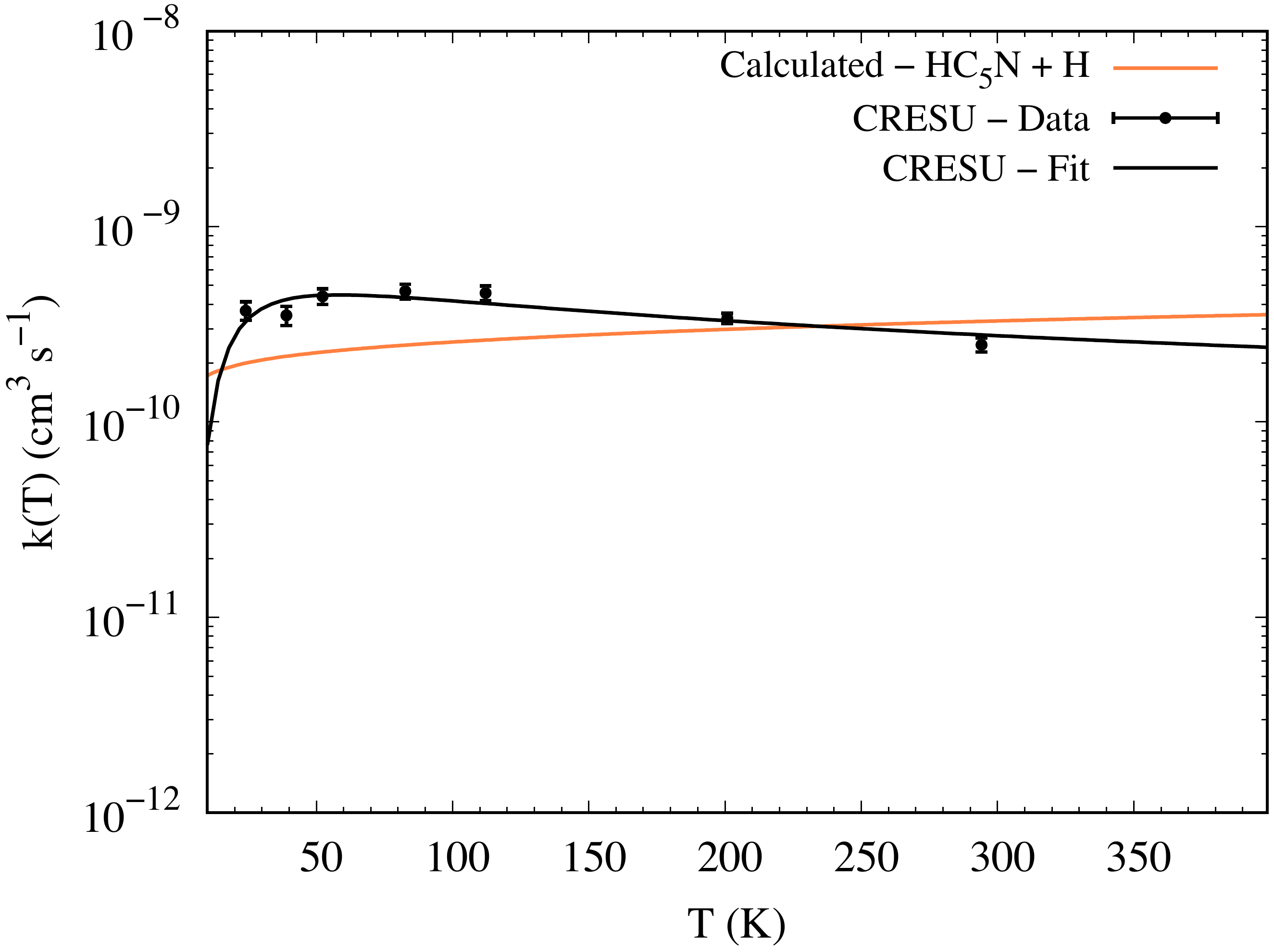}
    \caption{PES (\textit{left panel}) and rate coefficients (\textit{right panel}) of reaction [1] of Tab.~\ref{tab:revised-network}.
    The QM calculations are performed at the CCSD(T)/aug-cc-pVTZ//M06-2X/aug-cc-pVTZ level of theory. }
    \label{fig:pes-rate-reac-1}
\end{figure*}

\section{Detailed Theoretical Results} \label{sec:appendix-results-theo}

\subsection{Reaction [1]: \texorpdfstring{C${_3}$N + C$_{2}$H$_{2}$  $\rightarrow$  HC$_{5}$N + H}{reaction1} \label{subsec:append-theo-R1}}

\subsubsection{PES of reaction [1]}

The schematic PES for reaction [1] is shown in Fig.~\ref{fig:pes-rate-reac-1}. We identified six minima (MIN1, MIN2, MIN3, MIN4, MIN5 and ADD) and ten transition states (TS1, TS2, TS3, TS4, TS5, TS6, TS7, TS8, TS9 and TS-H). The reaction presents three exothermic channels leading to HC$_5$N + H ($\Delta H^0_0=-137.3$~\kjm), HC$_3$N + C$_2$H ($\Delta H^0_0=-14.7$~\kjm) and CN + C$_4$H$_2$ ($\Delta H^0_0=-53.4$~\kjm). The reaction can occur via two mechanisms: H-abstraction or C$_3$N addition to the $\pi$ bond of cyanoacetylene. Even if the hydrogen abstraction channel is exothermic by 14.7~\kjm, the presence of a barrier of 30.6~\kjm (TS-H) makes this channel not relevant at low temperatures. 
The addition of C$_3$N to cyanoacetylene gives rise, in a barrierless process, to an intermediate (MIN1) more stable than the reactants by 262.7~\kjm, in which the two hydrogen atoms assume a $trans$ configuration. MIN1 can easily isomerise to MIN2, overcoming a small barrier of 16.1~\kjm (represented by TS1), or to MIN3 by H migration. The isomerisation barrier is large, being 171.1~\kjm (represented by TS2). 
MIN2 is almost isoenergetic with MIN1 (-263.3~\kjm) since the only difference is in the $cis$ configuration of two hydrogen atoms, while MIN3 is more stable than MIN1 by 53.8~\kjm. Both MIN2 and MIN3 can dissociate into HC$_5$N + H by overcoming a barrier of 166.9~\kjm (TS3) and 192.7~\kjm (TS5), respectively. MIN2 can also isomerise to MIN4, located at -280.6~\kjm with respect to the reactants, following an H-migration process associated with an energy barrier of 192.3~\kjm (TS4). 
MIN4, in turn, can dissociate into HC$_5$N + H (exit barrier of 180.3~\kjm, TS7), or into HC$_3$N + C$_2$H (exit barrier of 262.9~\kjm, TS9). In the second case, the exit channel is characterized by the presence of an adduct ADD. MIN4 can also isomerise to MIN5 following the transfer of an H atom, a process that requires overcoming a barrier of 187.6~\kjm (TS6). From MIN5, the fission of the C-H bond leads to the formation of HC$_5$N + H by overcoming a barrier of 188.4~\kjm (TS8). Alternatively, the fission of the C-C bond leads to the formation of CN and C$_4$H$_2$ in a barrierless process.

\subsubsection{Kinetics of reaction [1]}

According to our calculations, the initial step of this reaction is the association of the two reactants with the formation of a C-C bond (MIN1). The alternative H-abstraction channel is a direct process (it cannot be included in the master equation) and is neglected in our calculations because of the presence of an activation energy of +30.8~\kjm. The possibility of back dissociation of MIN1 into the reactants is considered but found to be not particularly relevant because of the large stability of MIN1 (-262.7~\kjm). 
Since the reaction presents several exothermic channels, we calculated their branching fractions.
We found that over all the investigated temperature range (10-400 K) the main channel is the lone leading to HC$_5$N + H, with a branching fraction of $\sim1.0$. The other two channels leading to CN + C${_4}$H${_2}$ and HC${_3}$N + C${_2}$H are negligible. The HC$_5$N can be easily formed from MIN1 in a two-step process: isomerisation of MIN1 to MIN2 and a C-H bond fission. The other channels, instead, require more steps and are, in all cases, in competition, with HC$_5$N + H. Since this is the most exothermic channel, also MIN3, MIN4, and MIN5 preferentially form HC$_5$N + H. 
The calculated rate coefficient of the formation of cyclobutadiene was found to increase monotonically with temperature, following the trend of the capture coefficient (see Fig.~\ref{fig:pes-rate-reac-1}). The values of $\alpha$, $\beta$ and $\gamma$ derived from the fitting of the calculated rate coefficient for the temperature range 10-400~K are reported in Tab.~\ref{tab:revised-network}. The comparison of our results with the experimental data of \cite{fournier2014reactivity} and reported here (between 20 and 298 K) is shown in Fig.~\ref{fig:pes-rate-reac-1}. The results are consistent with the experimental points (same order of magnitude) but the temperature trend is not well reproduced. 
The $\alpha$, $\beta$, and $\gamma$ coefficients reported in \cite{fournier2014reactivity} are the result of the best fit of the experimental points. However, we note that this fitting only works in the 24-298~K temperature range, and it might not be realistic outside the range. This means that the values extrapolated to lower temperatures (10 K) might be incorrect and too low with respect to the real ones. 


\begin{figure*}
	\includegraphics[width=9.5cm]{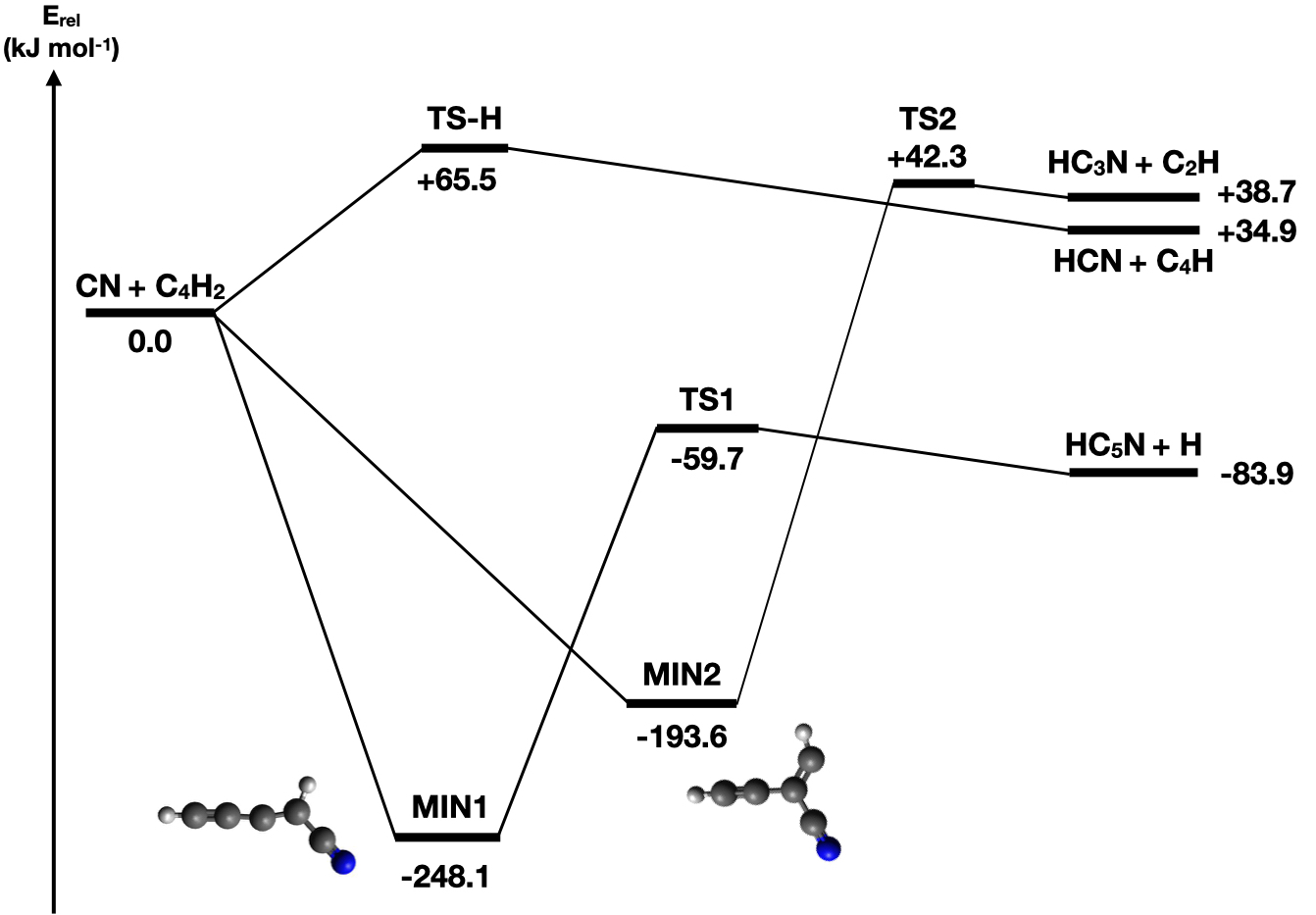}
 	\includegraphics[width=8cm]{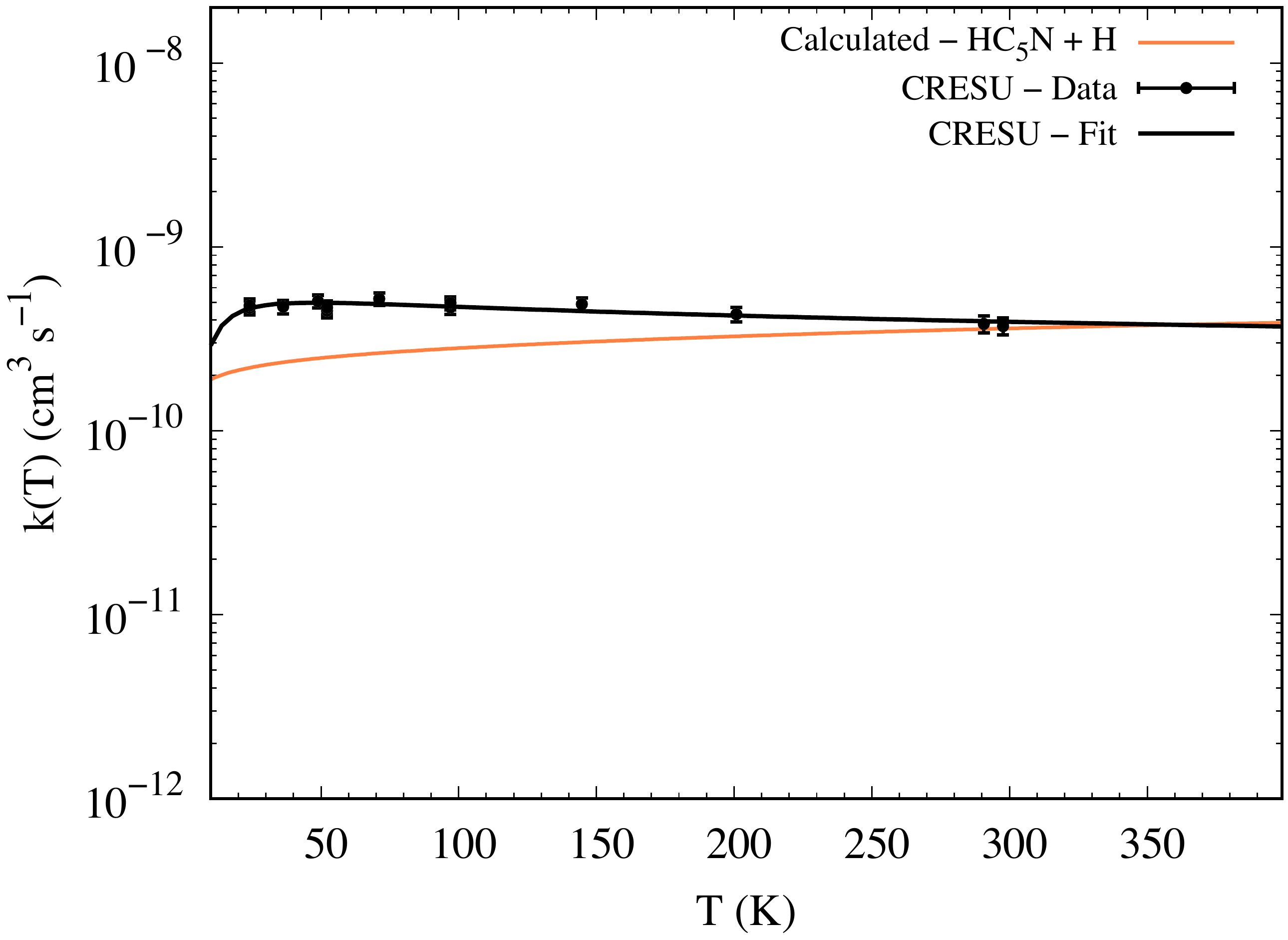}
    \caption{PES (\textit{left panel}) and rate coefficients (\textit{right panel}) of reaction [3] of Tab.~\ref{tab:revised-network}.
    The QM calculations are performed at the CCSD(T)/aug-cc-pVTZ//M06-2X/aug-cc-pVTZ level of theory. }
    \label{fig:pes-rate-reac-3}
\end{figure*}

\subsection{Reaction [3]: \texorpdfstring{C$_{4}$H$_{2}$ + CN $\rightarrow$ HC$_{5}$N + H}{reaction3} \label{subsec:append-theo-R3}}

\subsubsection{PES of reaction [3]}

The PES for reaction [3] is shown in Fig.~\ref{fig:pes-rate-reac-3}. Two attacks are possible: H-abstraction or CN addition to the $\pi$ bond of diacetylene. The H-abstraction pathway requires the system to overcome a barrier of 66.5~\kjm (represented by TS-H) and the HCN + C$_4$H products are formed with an enthalpy change of +34.9~\kjm. Due to the large endothermicity, the H-abstraction channel is not relevant under the typical conditions of the ISM.
We considered only the addition of the CN radical on the C-side since it has already been shown that the addition on the N side leads only to an endothermic channel \citep{zhang2009crossed}. Two different intermediates can be formed by the barrierless addition of CN on the triple bonds of C$_{4}$H$_{2}$: MIN1 (-248.1~\kjm with respect to the reactants) where CN is bonded to one of the terminal carbon atoms or MIN2 (-193.6~\kjm with respect to the reactants) where the CN is bonded to one of the central carbon atoms. We could not identify any vdW adduct in the entrance channel. The fission of a C-H bond in MIN1 leads to the formation of HC$_5$N + H (by overcoming a barrier of 188~\kjm, TS1), while the fission of a C-H bond in MIN2 leads to the formation of HC$_3$N + C$_2$H (by overcoming a barrier of 235.9~\kjm TS2). Therefore, TS2 is above the reactants energy asymptote, and the channel leading to HC$_3$N + C$_2$H is endothermic by 38.7~\kjm, 
As a consequence, the only relevant reaction channel is the one leading to HC$_5$N + H, which is exothermic by -83.9~\kjm.
The results are consistent with the experimental results reported here: the negative temperature dependence of the rate coefficient indicates a barrierless reaction \citep{cheikh2013}. 
The present results are in line with the previous theoretical studies \citep{fukuzawa1998neutral,tang2007dft,zhang2009crossed}. Note that \cite{tang2007dft} and \cite{zhang2009crossed} consider other possible attacks of the CN radical on ethylene and report more rearrangements of the intermediates. Nevertheless, also in those more complete studies, the only formed products are HC$_5$N + H.

\subsubsection{Kinetics of reaction [3]}

The calculated rate coefficients for the reaction [3] are shown in Fig.~\ref{fig:pes-rate-reac-3}. As for all the other reactions studied in this work, the H-abstraction channel was neglected, being characterized by a large entrance barrier 65.5~\kjm. The channel leading to HC$_3$N + C$_2$H is also not relevant because it is endothermic and TS2 lies 42.3~\kjm above the reactants. Thus, the only relevant channel at low temperature is the one leading to HC$_5$N + H, passing through MIN1 and TS1. The $\alpha$, $\beta$ and $\gamma$ parameters derived for the formation of HC$_5$N are reported in Tab.~\ref{tab:revised-network}. Our calculated rate coefficient is in good agreement with the experimental values (reported here and by \cite{cheikh2013}), but the experimental values at low temperatures are larger by a factor of $\sim$2 between 24~K and 100~K (see Tab.~\ref{tab:comp-exp-theo}). 


\begin{figure*}
	\includegraphics[width=9.2cm]{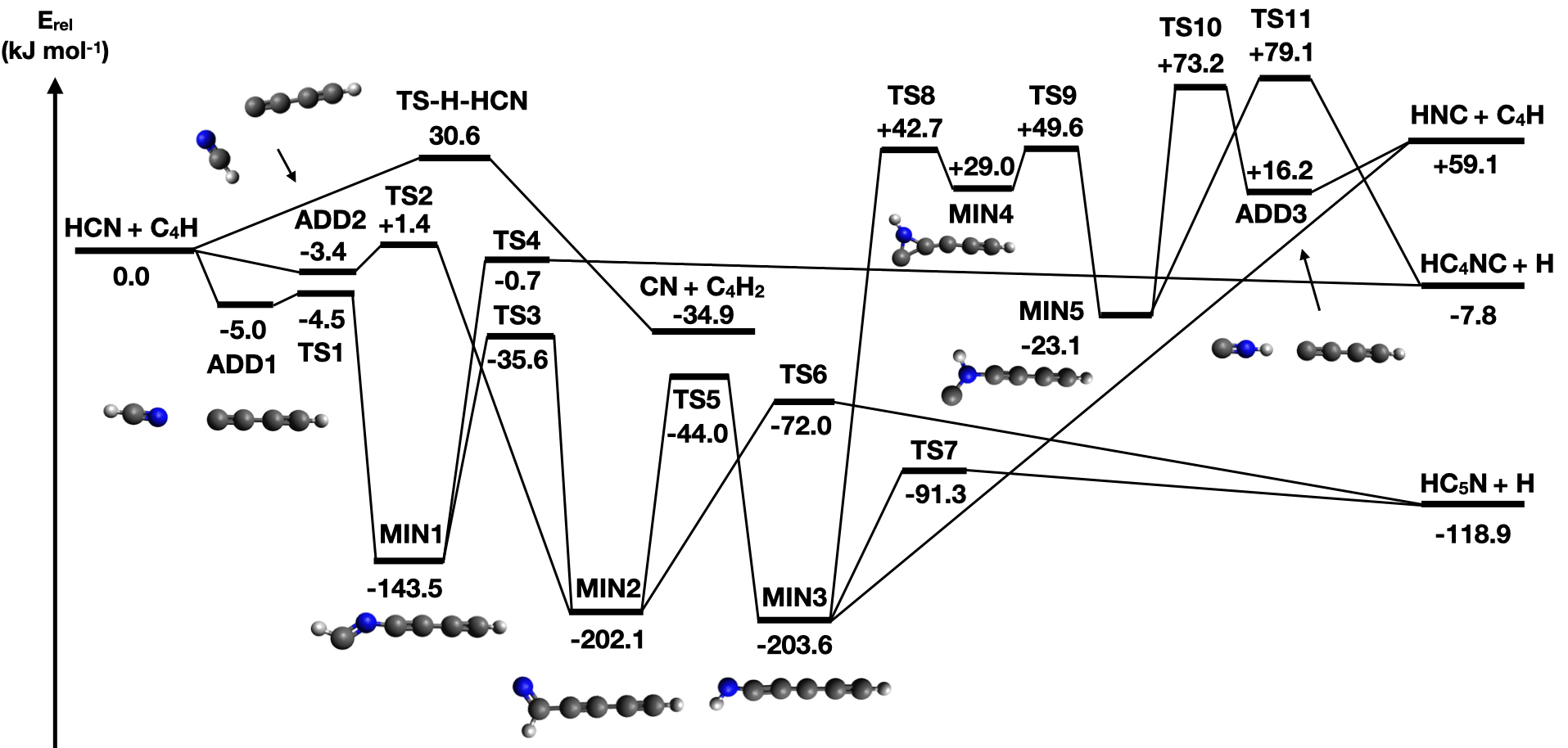}
 	\includegraphics[width=8.5cm]{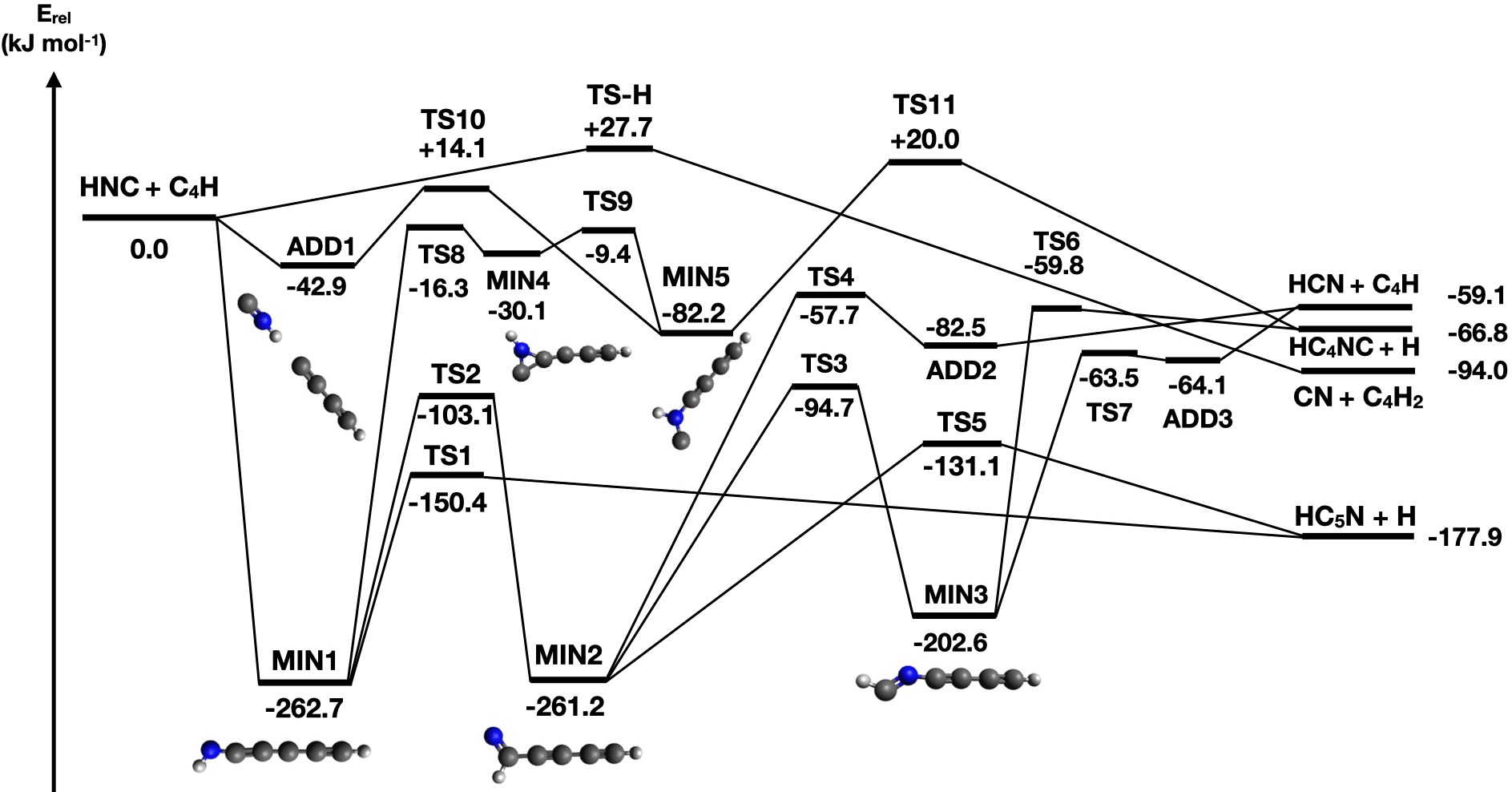}
    \caption{PESs of the reactions [4a] (\textit{left panel}) and [4b] (\textit{left panel}) of Tab.~\ref{tab:revised-network}. The QM calculations are performed at the CCSD(T)/aug-cc-pVTZ//M06-2X/aug-cc-pVTZ level of theory.}
    \label{fig:pes-reac4}
\end{figure*}

\subsection{Reactions [4a] and [4b]: \texorpdfstring{C$_{4}$H + HCN or HNC $\rightarrow$  HC$_{5}$N + H}{reaction4} \label{subsec:append-theo-R4}}

\subsubsection{PES of reaction [4a]}

The PES for the reaction [4a] is shown in Fig.~\ref{fig:pes-reac4}. We identified eight minima (ADD1, ADD2, ADD3, MIN1, MIN2, MIN3, MIN4 and MIN5) and twelve transition states (TS1, TS2, TS3, TS4, TS5, TS6, TS7, TS8, TS9, TS10, TS11 and TS-H). The reaction presents two possible mechanisms: H-abstraction mechanism (leading to CN and diacetylene, C$_4$H$_2$) of the C$_4$H addition to HCN. The H-abstraction channel is exothermic by 34.9~\kjm, but a barrier of 30.6~\kjm (TS-H) needs to be overcome. The addition of C$_4$H on HCN can occur in two different ways: the attack on the C-atom of HCN requires the formation of an initial adduct ADD2 which then converts into the covalently bound MIN2 (-202.1~\kjm) by overcoming a small barrier (TS2, +1.4~\kjm). Alternatively, the attack on the nitrogen atom of HCN leads to the formation of another adduct, ADD1, which converts into MIN1 (-143.5~\kjm) through TS1 (-4.5~\kjm). The two addition intermediates are connected through an isomerisation barrier (TS3) placed at -35.6~\kjm with respect to the reactants. 
MIN1 can directly dissociate into HC$_4$NC + H (this channel is slightly endothermic with a global enthalpy change of -7.8~\kjm) via TS4. 
MIN2 can dissociate into HC$_5$N + H by overcoming a barrier of 130.1~\kjm (TS6). MIN2 can also isomerise to MIN3, located at -203.6~\kjm via a barrier (TS5) of 158.1~\kjm. 
MIN3 can, in turn, dissociate into HC$_5$N + H by overcoming a barrier of 112.3~\kjm (TS7), or to HNC + C$_4$H in a barrierless process. MIN3 can also isomerise to a much less stable structure (+29.0~\kjm with respect to the reactants), MIN4, in which the nitrogen atom connects forms a  C-N-C cycle with the two last carbon atoms (a barrier of 246.3~\kjm, TS8, needs to be overcome). 
MIN4 can then convert into MIN5 by breaking the C-C bond of the C-N-C cycle, a process that requires overcoming a barrier of 20.6~\kjm (TS9). 
Finally, MIN5 can dissociate into HC$_4$NC + H ($\Delta H^0_0=-7.8$~\kjm) by overcoming a barrier of 96.3~\kjm (TS11), or  isomerise to ADD3 ($\Delta H^0_0=+16.2$~\kjm) by overcoming a barrier of 102.2~\kjm (TS10) and then dissociate into HNC + C$_4$H ($\Delta H^0_0=+59.1$~\kjm).

\subsubsection{Kinetics of reaction [4a]}

According to our calculations, the initial step of the reaction is the formation of ADD1 or ADD2.
Therefore, we calculated the capture rate coefficients for the formation of ADD1 and ADD2, solved the master equations for both channels and weighted the results according to the densities of states of the two vdW adducts. 
This is an approximate method to deal with the case of those reactions where more than one barrierless initial approach is possible (see, for instance, 
\citet{mancini2024unveiling}).

We found that, at all energies, the dominant channel is the one leading to HC$_5$N + H, while all the other channels are negligible. 
Because of the presence of transition states with energies comparable to the reactants (TS1 and TS2), the back dissociation is not negligible and causes the total rate coefficient to be lower than the capture rate coefficient. Nonetheless, the rate coefficient is found to be larger than the one assumed in the KIDA database, making this reaction possibly important for the formation of HC$_5$N. The $\alpha$, $\beta$ and $\gamma$ parameters derived for the formation of HC$_5$N are reported in Tab.~\ref{tab:revised-network}.



\subsubsection{PES of reaction [4b]}

The PES for the reaction [4b] is shown in Fig.~\ref{fig:pes-reac4}. We identified eight minima (ADD1, ADD2, ADD3, MIN1, MIN2, MIN3, MIN4 and MIN5) and twelve transition states (TS1, TS2, TS3, TS4, TS5, TS6, TS7, TS8, TS9, TS10, TS11 and TS-H). As for reaction [4a], the reaction [4b] presents an H-abstraction mechanism to form CN and diacetylene C$_4$H$_2$. However, even if the channel is exothermic (the enthalpy change for this channel is -94.0~\kjm), the presence of an activation energy of 27.7~\kjm (TS-H) makes it not relevant.
The alternative mechanism is the addition of C$_{4}$H to HNC, which can occur in two different ways, as for reaction [4a]. The addition can indeed form two intermediates: MIN1 if C$_{4}$H adds on the C side or MIN5 if C$_{4}$H adds on the N side (in this case, a vdw adduct, ADD1, is formed and a small barrier, TS10, needs to be overcome). 
MIN1 (located at -262.7 ~\kjm) is much more stable than ADD1 (-42.9~\kjm) and MIN5 (-82.2~\kjm), respectively. The formation of MIN1 is a barrierless process (we could not identify a vdW adduct in this case).
Both MIN5 and MIN1 can lose an H atom and form HC$_4$NC + H ($\Delta H^0_0=-118.9$~\kjm) and HC$_5$N + H ($\Delta H^0_0=-150.4$~\kjm), respectively. However, the formation of HC$_4$NC + H requires overcoming a barrier (TS11, + 20.0~\kjm) above the reactants, while the formation of HC$_5$N + H from MIN1 requires passage through TS1 which is at -150.4~\kjm. MIN1 can also isomerise to MIN2 (-261.2~\kjm) by overcoming a barrier of 159.6~\kjm (TS2). 
The two addition mechanisms that form MIN1 and MIN5 are connected one to the other through MIN4, an intermediate at -30.1~\kjm with respect to the reactants, which presents a C-N-C cyclic structure. MIN5 can indeed isomerise to MIN4 through TS9 (-9.4~\kjm), and MIN4 can isomerise to MIN1 through TS8 (-16.3~\kjm).
MIN2, in turn, can dissociate into HC$_5$N + H by overcoming a barrier of 130.1~\kjm (TS5). Alternatively, it can isomerise to MIN3 by overcoming a barrier of 166.5~\kjm (TS3) or ADD2 by overcoming a barrier of 203.5~\kjm (TS4). Finally, MIN3 can dissociate into HC$_4$NC + H by overcoming a barrier of 142.8~\kjm (TS6), or it can isomerise to ADD3 by overcoming a barrier of 139.1~\kjm (TS7). Both ADD2 and ADD3 can then dissociate into C$_{4}$H + HCN, with a global enthalpy change of -59.1~\kjm.

\begin{figure*}
	\includegraphics[width=9cm]{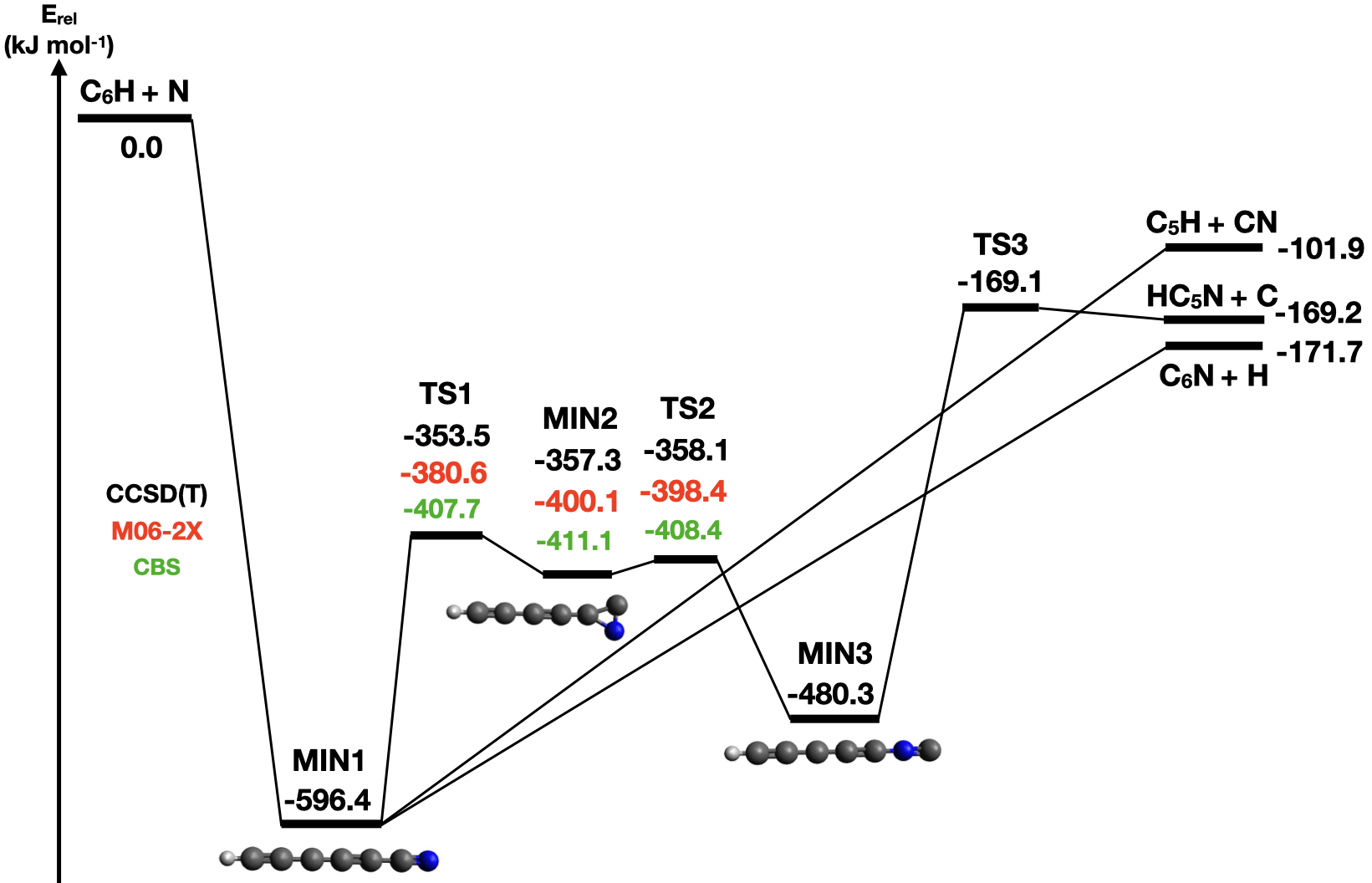}
 	\includegraphics[width=8cm]{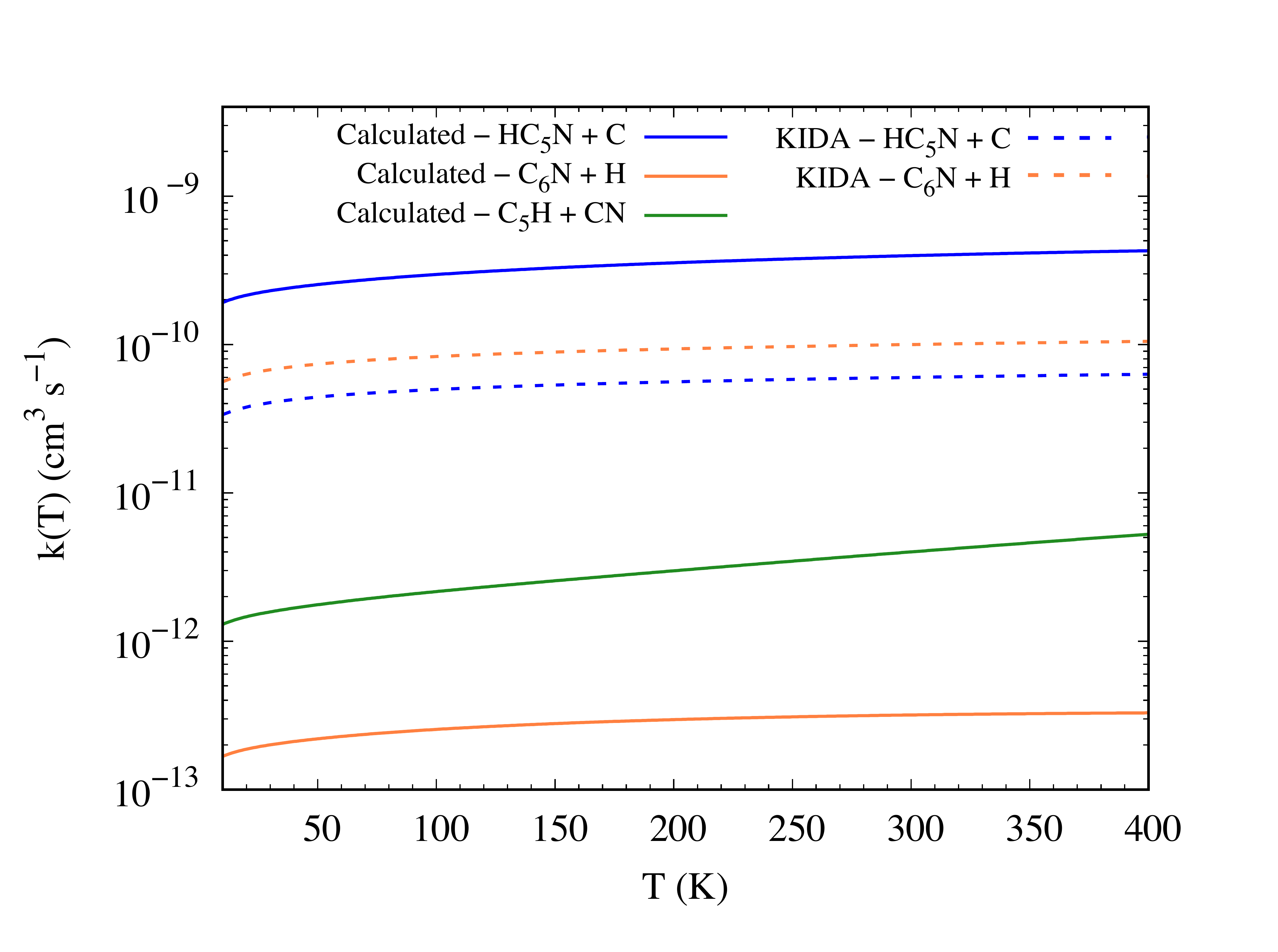}
    \caption{Left panel: PES of the reaction [5] of Tab.~\ref{tab:revised-network}.
    The QM calculations are performed at the CCSD(T)/aug-cc-pVTZ//M06-2X/aug-cc-pVTZ level of theory. Right panel: rate coefficients for the formation of C$_6$N + H (orange), HC$_5$N + C (blue) and C$_5$H + CN (green) of reaction [5] of Tab.~\ref{tab:revised-network}. Solid lines correspond to rate coefficients calculated using the methodology described in Sec.~\ref{subsubsec:theo-methods-kinetics}, while dashed lines correspond to rate coefficients reported in the KIDA database \citep{wakelam2012kinetic,loison2014interstellar} }
    \label{fig:pes-rate-reac-5}
\end{figure*}

\subsubsection{Kinetics of reaction [4b]}


For this reaction, the only relevant channel is the one leading to HC$_5$N + H via the formation of MIN1. We characterized the entrance channel and found a capture rate coefficient of about $2\times10^{-10}$  cm$^3$ s$^{-1}$. Since the first intermediate MIN1 is formed with a large energy gain (-262.7~\kjm), the back-dissociation is negligible and the rate has almost no dependence on the temperature. HC$_5$N is the main product of the reaction, with a branching fraction of $\sim$1.0 in the investigated temperature range (10-400 K). The $\alpha$, $\beta$ and $\gamma$ parameters derived for the formation of HC$_5$N are reported in Tab.~\ref{tab:revised-network}.


\subsection{Reaction [5]: \texorpdfstring{C${_6}$H + N $\rightarrow$ HC$_{5}$N + C}{reaction5} \label{subsec:append-theo-R5}}

\subsubsection{PES of reaction [5]}

The PES for reaction [5] is shown in Fig.~\ref{fig:pes-rate-reac-5}. We identified three minima (MIN1, MIN2 and MINI3) and three transition states (TS1, TS2 and TS3). The barrierless addition of N ($^4$S) on the terminal C atom of C$_6$H ($^2\Sigma$) leads to the formation of the first intermediate MIN1 in a triplet state, which is 596.4~\kjm more stable than the reactants. MIN1 can directly dissociate into C$_6$N + H ($\Delta H^0_0=-171.7$~\kjm) or C$_5$H  + CN ($\Delta H^0_0=-101.9$~\kjm) in barrierless processes, following the fission of a C-H or C-C bond, respectively. MIN1 can also isomerise to MIN2, placed at -357.3~\kjm with respect to the reactants at CCSD(T)/aug-cc-pVTZ level, by overcoming a barrier of 242.9~\kjm (TS1). MIN2 easily isomerises into MIN3 through a transition state TS2 located above MIN2 at M06-2X/aug-cc-pVTZ level (+1.7~\kjm with respect to MIN2), but not at CCSD(T)/aug-cc-pVTZ level (-0.8~\kjm with respect to MIN2). 
By using the CBS extrapolation for TS1, MIN2, and TS2, we confirmed that both TS1 and TS2 lie above the energy of MIN2. 
MIN3 (-480.3~\kjm with respect to the reactants) can then form HC$_5$N + C ($\Delta H^0_0=-169.2$~\kjm) following the fission of the terminal C-N bond. A barrier of 311.2~\kjm (TS3) needs to be overcome. 

These results are consistent with the results reported for the reaction of N atoms with C$_2$H, obtained with a similar level of theory (M06-2X//aug-cc-pVQZ) by \cite{loison2015ab}. In both cases, the reaction presents a barrierless addition leading to a linear intermediate HC$_6$N with symmetry C$_{\infty v}$, in which the nitrogen atom is bonded to the terminal carbon atom. From this intermediate, the nitrogen atom can migrate, forming an intermediate with a cyclic C-N-C structure and then an intermediate HC$_4$NC in which the positions of the terminal C and N atoms are inverted. 



\renewcommand{\arraystretch}{1.8}

\begin{table*}
    \begin{center}
    \caption{Comparison of theoretical and experimental total rate coefficients for the reactions [1], [2], and [3]. We report the values of $\alpha$, $\beta$ and $\gamma$ and the values of $k$ at 24 and 100 K. The experimental values are those reported in Section \ref{sec:experiments-results} of the current work, and correspond to those in \protect\cite{cheikh2013} and \protect\cite{fournier2014reactivity}. The theoretical values, obtained using the procedure presented in Sec.~\ref{subsubsec:theo-methods-kinetics}, are described in details in Sec.~\ref{subsec:append-theo-R1},\ref{subsubsec:theo-results-reaction2} and \ref{subsec:append-theo-R3}.} 
    \label{tab:comp-exp-theo}
	\begin{tabular}{cl||ccccc|ccccc}
		\hline
\multicolumn{2}{c||}{Reaction} &  \multicolumn{5}{c|}{Experimental} & \multicolumn{5}{c}{Calculated}  \\
\multicolumn{2}{c||}{} & $\alpha$ & $\beta$ & $\gamma$ & k (24~K) & k (100~K) & $\alpha$ & $\beta$ & $\gamma$  & k (24~K) & k (100~K) \\
	\hline
		\hline
1            & C$_{3}$N + C$_2$H$_{2}$       &  3.09$\times10^{-10}$  & -0.577 & -33.64   & 2.5$\times10^{-10}$  &  4.2$\times10^{-10}$  &  3.27$\times10^{-10}$  & 0.24   & -0.43  & 2.0$\times10^{-10}$  & 2.6$\times10^{-10}$ \\
2$^{\rm a}$  & HC$_{3}$N + C$_2$H            &  3.91$\times10^{-11}$  & -1.04  & 0.0      & 5.6$\times10^{-10}$  &  1.2$\times10^{-10}$  &  6.56$\times10^{-11}$  & -0.26  &  2.33  & 1.0$\times10^{-10}$  &  8.5$\times10^{-11}$\\
3            & C$_4$H$_{2}$ + CN             &  4.06$\times10^{-10}$  & -0.24  & 11.5     & 4.6$\times10^{-10}$  &  4.7$\times10^{-10}$  &  3.58$\times10^{-10}$  & 0.20   & 0.0    & 2.2$\times10^{-10}$  & 2.8$\times10^{-10}$ \\
    \hline
	\end{tabular}\\
 $^{\rm a}$ The $\alpha$, $\beta$ and $\gamma$ coefficients derived from theoretical calculations of this reaction are valid only in the 10--150~K temperature range.\\  
	\end{center}
\end{table*}

\subsubsection{Kinetics of reaction [5]}

The calculated rate coefficients for the reaction [5] are shown in Fig.~\ref{fig:pes-rate-reac-5}. 
Because of some difficulties in carrying out variational calculations for the barrierless channel leading to C$_6$N + H, we have assumed as transition states for the system the products at infinite separation. 
Regarding the energy of TS2, for the unimolecular calculations we have used the energy calculated with CBS extrapolation.
Back-dissociation is found to be negligible since the initially formed minimum (MIN10) has a well of -596.4~\kjm. As a result, rate coefficients are essentially independent of temperature, with small values of $\beta$ and $\gamma=0$ (see Tab.~\ref{tab:revised-network}). 
%
%
The channel leading to HC$_5$N is found to be the dominant one, with a BF of 0.992. The other two channels (C$_6$N + H and C$_5$H + CN) give a small contribution, with BFs of 0.007 and 0.001, respectively.
The comparison of our results with the values found by \cite{loison2014interstellar} and reported in KIDA \citep{wakelam2012kinetic} is also shown in Fig.~\ref{fig:pes-rate-reac-5}. Our calculated total rate coefficient for the reaction, which corresponds approximately to the capture coefficient, is consistent with the rate calculated by \cite{loison2014interstellar}. However, the branching fractions are found to have an opposite trend, with HC$_5$N + C being the main products instead of C$_6$N + H. The exothermicity of the two channels is comparable, and C$_6$N + H was expected to be favored because of the fewer rearrangements needed. However, the unimolecular rate for the formation of C$_6$N + H from MIN1 is an order of magnitude smaller than the unimolecular rate for the rearrangement of MIN1 into MIN2, making the formation of HC$_5$N more efficient.




%

\section{Astrochemical models} \label{sec:appendix-models}

\begin{table}
    \begin{center}
	\caption{Initial elemental abundances relative to H nuclei adopted for the cold prestellar core modeling. The abundances are adapted from \citep{jenkins2009unified}.}
	\label{tab:astro-cold-initial-abd}
	\begin{tabular}{cc|cc}
    \hline
    Element & Abundance & Element & Abundance \\
    \hline
    \hline
He      & $9.0 \times 10^{-2}$  &  P$^+$   & $2.0 \times 10^{-10}$ \\
C$^+$   & $2.0 \times 10^{-5}$  &  Na$^+$  & $2.0 \times 10^{-9}$\\
O       & $2.6 \times 10^{-5}$  &  Fe$^+$  & $3.0 \times 10^{-9}$ \\
N       & $6.2 \times 10^{-6}$  &  Cl$^+$  & $1.0 \times 10^{-9}$\\
S$^+$   & $8.0 \times 10^{-8}$  &  F$^+$   & $1.0 \times 10^{-9}$\\
Si$^+$  & $8.0 \times 10^{-9}$  &  & \\
	\hline
	\end{tabular}
	\end{center}
\end{table}

\begin{table}
    \centering
        \caption{Parameters used to model the passage of the shock (see Sec.~\ref{subsec:astro-model-shock}).
    The upper half table lists the adopted H$_{2}$ density, n$_{H_2}$, temperature, $T$, and cosmic-ray ionisation rate, $\zeta_{CR}$, of the gas after the passage of the shock.
    The lower half table lists the adopted abundance of the species injected into the gas-phase from the grain mantle after the passage of the shock.
    }
    \begin{tabular}{c|c}
        \hline
        \hline
        \multicolumn{2}{c}{Physical parameters of the shocked gas} \\
        Parameter & L1157-B1 value \\
        \hline
        n$_{H_2}$    [cm$^{-3}$] & $4\times 10^5$      \\
        T            [K]         & 90                  \\
        $\zeta_{CR}$ [s$^{-1}$]  & $6\times 10^{-16}$ \\
        \hline
        \hline
        \multicolumn{2}{c}{Abundances (wrt H) of the injected species} \\
        Species & L1157-B1 value \\
        \hline
        H$_{2}$O    & $1\times 10^{-4}$ \\
        CO$_{2}$    & $3\times 10^{-5}$ \\
        CO    & $1\times 10^{-4}$ \\
        CH$_3$OH  & $4\times 10^{-6}$ \\
        NH$_3$    & $5.6\times 10^{-5}$ \\
        H$_{2}$CO   & $1\times 10^{-6}$ \\
        OCS       & $2\times 10^{-6}$ \\
        SiO       & $1\times 10^{-6}$ \\
        Si        & $1\times 10^{-6}$\\
        CH$_3$CH$_{2}$ & $4\times 10^{-8}$  \\
        CH$_3$CH$_{2}$OH & $6\times 10^{-8}$ \\
        SiH$_4$  & $1\times 10^{-7}$ \\
        \hline
    \end{tabular}
    \label{tab:model-parameters-shock}
\end{table}

\bsp	
\label{lastpage}
\end{document}